\documentclass[prd,twocolumn,showpacs,preprintnumbers,amsmath,amssymb]{revtex4}

\usepackage{graphicx}
\usepackage{dcolumn}
\usepackage{bm}
\usepackage{wrapfig} 
\usepackage{multirow}
\usepackage{color}

\newcommand{\beq}{\begin{eqnarray}} 
\newcommand{\eeq}{\end{eqnarray}}

\begin{document}

\preprint{CERN--PH--TH/2015--114}\vspace*{0.5cm}

\title{LHC constraints on Gravitino Dark Matter}

\author{Alexandre Arbey}
\email{alexandre.arbey@ens-lyon.fr}
\affiliation{Centre de Recherche Astrophysique de Lyon, Observatoire de Lyon,\\
 Saint-Genis Laval Cedex, F-69561; CNRS, UMR 5574;\\ 
 Ecole Normale Sup\'erieure de Lyon;\\
 Universit\'e de Lyon, Universit\'e Lyon 1, F-69622~Villeurbanne Cedex, France\\ and CERN, CH-1211 Geneva, Switzerland}%
\author{Marco Battaglia}
\email{Marco.Battaglia@ucsc.edu}
\affiliation{Santa Cruz Institute of Particle Physics,\\
  University of California at Santa Cruz, CA 95064, USA\\
  and CERN, CH-1211 Geneva, Switzerland}%
\author{Laura Covi}
\email{laura.covi@theorie.physik.uni-goettingen.de}
\affiliation{Institute for theoretical Physics, Georg-August-Universit\"at G\"ottingen, D-37077 G\"ottingen, Germany}%
\author{Jasper Hasenkamp}
\email{jasper.hasenkamp@nyu.edu}
\affiliation{Center for Cosmology and Particle Physics, New York University, NY 10003, USA}%
\author{Farvah Mahmoudi}
\email{mahmoudi@in2p3.fr}
\affiliation{Centre de Recherche Astrophysique de Lyon, Observatoire de Lyon,\\
  Saint-Genis Laval Cedex, F-69561; CNRS, UMR 5574;\\ 
  Universit\'e de Lyon, Universit\'e Lyon 1, F-69622~Villeurbanne Cedex, France\\ and CERN, CH-1211 Geneva, Switzerland}%

\begin{abstract}
  Gravitino Dark Matter represents a compelling scenario in Supersymmetry, which brings together
a variety of data from cosmology and collider physics. We discuss the constraints obtained 
from the LHC on supersymmetric models with gravitino dark matter and neutralino NLSP,
which is the case most difficult to disentangle at colliders from a neutralino LSP forming DM.
The phenomenological SUSY model with 19+1 free parameters is adopted.
Results are obtained from broad scans of the phase space of these uncorrelated parameters. 
The relation between gravitino mass, gluino mass and reheating temperature as well as the
derived constraints on these parameters are discussed in detail. 
This relation offers a unique opportunity to place stringent bounds on the cosmological model, 
within the gravitino dark matter scenario, from the results of the LHC searches in Run-2 and the planned 
High-Luminosity upgrade.
\end{abstract}

\pacs{11.30.Pb, 14.80.Ly, 14.80.Nb, 95.35.+d}

\maketitle

\section{Introduction}
\label{sec:intro}

With the discovery of the Higgs boson at the LHC~\cite{ATLAS:2012zz,CMS:2012zz},
collider particle physics moves on to a new phase, 
where searches for new phenomena will be at the center of the physics program. Dark matter (DM) plays a
special role in connecting these searches to cosmology. Not only DM
appears to be the most solid indication of new physics obtained so far, but the precision of the determination
of its relic density in the Universe, through the study of the cosmic microwave background (CMB)~\cite{Ade:2013zuv},
and the increasing sensitivity of the searches for DM scattering in underground
experiments~\cite{Ahmed:2009zw,Aprile:2012nq,Akerib:2013tjd} represent powerful
constraints for most models of new physics that offer a DM candidate.

Supersymmetry (SUSY) is one of the most promising and best motivated extensions of the Standard Model
of particle physics (SM), providing at the same time a solution to the hierarchy problem 
within a weakly coupled theory, radiative electroweak symmetry breaking with a light Higgs
scalar with properties that, in the decoupling limit, come to resemble those of the Standard Model,
and also naturally includes several viable dark matter candidates within its extensive particle
spectrum.

Searches conducted at the LHC on the 7+8~TeV Run-1 data have already set some significant constraints on
low energy SUSY, in particular on the mass of the coloured superpartners and on the general mass scale
in the case of highly constrained models, such as the CMSSM and the
NUHM models~\cite{Bechtle:2013mda, Buchmueller:2013rsa, Roszkowski:2014wqa}.
However, many scenarios with supersymmetric particles within the reach of the LHC runs at 13-14~TeV are totally
unprobed at present, in particular when we consider more general SUSY
models~\cite{Djouadi:1998di,Henrot-Versille:2013yma, Craig:2013cxa,Nath:2015dza} such as the ``phenomenological MSSM'' 
(pMSSM), without any assumption on the SUSY breaking mediation mechanism and on unification at high scale.
This model gives the most general framework in the minimal SUSY extension of the SM (MSSM)
for studying supersymmetric signals and constraints from colliders. It is being adopted by an increasing
number of phenomenological~\cite{AbdusSalam:2009qd,Conley:2011nn, Arbey:2011un, Sekmen:2011cz, Arbey:2011aa, CahillRowley:2012cb, CahillRowley:2012rv, Arbey:2012dq, CahillRowley:2012kx,
Arbey:2012bp, Arbey:2013iza, Kowalska:2014hza} and experimental~\cite{CMS:2013rda,CMS:2014mia} studies.

The particle most studied in conjunction with SUSY dark matter is the lightest neutralino, $\tilde{\chi}^0_1$, 
which provides us with a viable realisation of the WIMP mechanism and implies DM signals not only at colliders,
but also in direct and indirect detection. However, another very well-motivated SUSY candidate
for DM is the gravitino, $\tilde{G}$, the superpartner of the graviton which couples with
Planck-suppressed strength to the rest of the model, making it very difficult to detect in direct or indirect
DM experiments, in SUSY models with conserved R-parity. Even a small R-parity breaking opens up the
possibility of gravitino decaying DM candidate relaxing part of the cosmological constraints~\cite{Buchmuller:2007ui}. 
Colliders offer a unique way to probe the DM sector for scenarios with gravitino dark matter.
Therefore, dedicated searches should be actively pursued to fully exploit the LHC reach in
these scenarios.

In general, assuming that the gravitino is the Lightest SUSY Particle (LSP) and forms DM greatly relaxes the
DM constraints on the SUSY spectrum, since its abundance is not produced through the WIMP mechanism and does not only
depend on the SUSY spectrum, but also on the reheating temperature in the Universe after
inflation~\cite{Bolz:2000fu,Pradler:2006qh}.
Consequently, the $\tilde{G}$ remains a good candidate for DM even if the spectrum of the superpartners turns out
to be much heavier than expected. Nevertheless, we will show that the correlation between the gluino (and in general gaugino)
mass and gravitino production ensures that the LHC will be able to probe the gravitino production mechanism
for high reheating temperatures above $10^9$~GeV and that the High Luminosity LHC program (HL-LHC) may play a crucial
role for these tests. 
 
In this paper we present a study of the present and future constraints on SUSY models
with gravitino LSP responsible for DM and neutralino Next-to-Lightest SUSY Particle (NLSP). These models
are not yet strongly constrained since the current LHC limits on weakly-produced SUSY particles are limited in coverage
and our scenarios of interest not easy to disentangle from those of neutralino LSP. Indeed, here we consider the parameter
space where the neutralino is stable on collider time scales and gives rise to the usual missing transverse energy (MET)
signatures at the LHC.
After considering the present bounds on the model and how strongly the cosmological and astrophysical constraints
limit the parameter space, we assess the capability of the forthcoming LHC runs to test the gravitino
production mechanism and to tell the gravitino LSP from the neutralino LSP solutions.

Compared to previous studies, this analysis implements the latest collider constraints, including those from
monojet searches which complement the other searches in the regions of the parameter space with degenerate
SUSY masses, and we discuss the combination of LHC and other data in the context of the gravitino models.
We adopt the pMSSM as a generic MSSM model that, contrary to more constrained models used in the
past~\cite{Ellis:2003dn,Roszkowski:2004jd,Cerdeno:2005eu,Cyburt:2006uv,Choi:2007rh,Ellis:2008as,Bailly:2009pe}
for similar studies, does not imply relations between the masses of the different SUSY particles. This opens up
new region of the parameter space with interesting phenomenology and search opportunities at the LHC which were not
available in the constrained MSSM (CMSSM), as we demonstrate in this study.
The concurrent analysis of the LHC data and the DM direct detection experimental results will
be key to identify a neutralino NLSP signal without observation of the decay or to exclude a large fraction of the model
parameter space. The combination of the LHC sensitivity on the gluino mass and the requirement of a large enough
reheating temperature after inflation for thermal leptogenesis~\cite{Giudice:2003jh,Fujii:2003nr,Buchmuller:2004nz,Fong:2013wr}
is specific to the study performed in the pMSSM, restricts the viable MSSM parameter space and highlights the capability
of the LHC to test these cosmological scenarios.

This paper is organised as follows: after discussing the generalities of the gravitino LSP and DM scenario
together with the cosmological bounds from nucleosynthesis in chapter II, we describe our scan strategy for
the pMSSM parameter space and the collider and low-energy constraints in chapter III. Chapter IV presents the
results of our analysis, while in chapter V we give our conclusions.

\section{Gravitino Dark Matter}

The gravitino is the superpartner of the graviton in models with local supersymmetry~\cite{Wess:1992cp}.
It is the particle most directly related to the effect of SUSY breaking and obtains its mass via 
the SuperHiggs mechanism \cite{Binetruy:2006ad} as the {\it gauge-fermion} of SUSY from any existing 
source of breaking.
The hierarchy between the masses of the superpartners is determined by the particular
mediation mechanism, and it can result in a gravitino which is heavier, lighter or even much
lighter than the other superpartners. If the gravitino is the lightest SUSY state and therefore absolutely
(or sufficiently) stable, it represents a viable candidate for explaining dark matter. 
The gravitino couplings are set by supergravity and by the MSSM parameters and are 
inversely proportional to the reduced Planck mass, taken here at its standard value of $ M_P = 2.4 \times 10^{18} $ GeV.
The gravitino mass is the only additional parameter needed to describe the gravitino and its interactions.
Starting with the phenomenological MSSM with 19 parameters set at the electroweak scale, we will add $M_{\tilde{G}}$
as independent parameter, bringing the total to 20. 

In this paper, we shall focus our analysis on the part of the parameter space where the gravitino is
the LSP and a cold DM candidate. We do not consider a very light gravitino, with mass in the eV to few keV range,
since it is at odds with present cosmological data which strongly disfavour hot or warm Dark Matter and may 
only constitute a subdominant DM component~\cite{Viel:2005qj, Boyarsky:2008xj,Feng:2010ij,Markovic:2013iza}.
In this region of parameters, another (cold) DM candidate is needed and no other supersymmetric particle can
fulfill this role. Moreover, a very light gravitino can be a warm thermal relic only in presence of entropy
production with a sufficient dilution factor of the gravitino freeze-out density~\cite{Baltz:2001rq}.  
Still, this scenario has been widely studied, in particularly in relation to the gauge-mediated
SUSY breaking (GMSB) mechanism~\cite{Dimopoulos:1996vz,Giudice:1998bp}. It gives very distinctive signatures at
colliders since the NLSP decays promptly leading to photons/jet + MET, which are actively searched by the ATLAS and
CMS experiments at the LHC~\cite{deAquino:2012ru, CMS-PAS-SUS-12-018,ATLAS-CONF-2014-001, Maltoni:2015twa}.
In the scenarios we explore here, the Next-to-Lightest SUSY Particle (NLSP) often mimicks a stable LSP,  even if
the NLSP can be in principle charged under the SM group and therefore cannot be stable on cosmological scales. 
We assume the NLSP to be the lightest neutralino, $\tilde{\chi}^0_1$, which is the case most difficult to disentangle 
at colliders from the scenario of neutralino LSP and Dark Matter particle. 

Although very weakly coupled, gravitinos can be produced in substantial numbers
in the primordial thermal bath from scattering of SUSY particles, in particular
the colored or electroweakly charged states. Usually the interactions involving gluinos or gauginos
gives the dominant contribution since they are some of the most abundant and strongly coupled
particles and interact with gravitinos by an effective dimension-5 operator, leading to a yield
proportional to the highest equilibrium temperature of the thermal bath~\cite{Bolz:2000fu,Pradler:2006qh}.
For a light gravitino, this contribution is dominated by the Goldstino component of the gravitino and given by
\begin{equation}
\Omega_{\tilde{G}}^{th}\,h^2=0.83\; \frac{T_{RH}}{10^9\,\mbox{GeV}} 
\left(\frac{m_{3/2}}{1\,\mbox{GeV}}\right)^{-1}\sum_{i=1}^3 \gamma_{i} (T_{RH})
\left(\frac{M_{i}}{300\,\mbox{GeV} }\right)^2 
\label{eq:thermalprod}
\end{equation}
where $ \gamma_{i} (T_{RH}) $ are numerical factors including the dependence on the gauge coupling
for each SM gauge group and the evolution of the gauge couplings and gaugino masses $ M_{i} $ to the 
scale of the reheating temperature $T_{RH}$. To do so, we used here only one-loop RGEs to capture the
main effect and we define, following~\cite{Pradler:2006qh},
\begin{equation}
\gamma_i (T) =  y_i \frac{g_i^6(T)}{g_i^4(M_i)}  \log \left[\frac{k_i}{g_i(T)} \right]
\end{equation}
 with $(y_i, k_i)$ equal to $(4.276, 1.271)$, $(1.604, 1.312)$, $(0.653, 1.266)$ for the SM gauge groups 
 $ SU(3)$, $SU(2)$ and $U(1)$, respectively. It is clear from these factors that the
 gluinos often give the largest contribution, due to their multiplicity and their larger gauge coupling.
 The expression given in eq.~(\ref{eq:thermalprod}) is valid only in the perturbative regime, and becomes
 less reliable  at low reheat temperatures, where $ g_3 $ becomes large. Note though, that for low $ T_{RH} $,
 in the region of supersymmetric masses we will explore, gaugino scatterings do not provide the dominant 
 contribution to the gravitino relic density and therefore an order one determination of this part is 
 sufficient for our purposes.
  
Processes involving scalar SUSY particles, like squarks and sleptons, include also dimension-4 type couplings,
which generate a yield independent of the thermal bath temperature, but those contributions are often negligible,
unless a strong hierarchy between scalar and gaugino masses is assumed~\cite{Hall:2009bx,Cheung:2011nn}.
The contribution from the NLSP decay after freeze-out can be quite substantial, especially for the case of a
Bino-like neutralino NLSP. Such contribution is given by~\cite{Covi:1999ty,Feng:2003xh, Feng:2003uy}
\begin{equation}
\Omega_{\tilde{G}}^{SW}\,h^2 =  \left(\frac{m_{3/2}}{M_\chi }\right) \Omega_{\chi} \,h^2\; ,
\label{eq:outeqprod}
\end{equation}
where $ M_\chi $ is the mass of the lightest neutralino and $ \Omega_{\chi} \,h^2 $ its present 
energy density, if the neutralino would be stable, set by the freeze-out process. 

In determining the relic gravitino density we consider the thermal contribution from gaugino scattering
given in eq.~(\ref{eq:thermalprod}) as well as that arising from neutralino decay out of equilibrium
in eq.~(\ref{eq:outeqprod}). 
For any value of the pMSSM parameters and gravitino mass, a specific value of the
reheat temperature allows us to match the observed DM abundance, as long as the contribution
from NLSP decay is smaller than the observed DM density. In the pMSSM parameter region
where the neutralino relic density exceeds the DM energy density, an upper bound on the gravitino mass also
appears. This bound may be stronger than our request that the gravitino is the LSP.  In general, there is a
maximal value of the reheating temperature allowed at any parameter point, and possibly also a maximal gravitino
mass, in order to avoid gravitino density overclosure.

The total gravitino density can be written in a simplified form as
\begin{equation}
\Omega_{\tilde{G}} h^2 = A\; \frac{T_{RH}}{m_{3/2}} + B\; m_{3/2}
\label{eq:totalgravdensity}
\end{equation}
where the parameters $A$ and $B$ are functions of the pMSSM parameters, as can be read from
eqs.~(\ref{eq:thermalprod}),(\ref{eq:outeqprod}). $A$ depends mostly on the gaugino masses
$M_1$, $M_2$ and $M_3$, while $B$ contains the neutralino relic density and depends directly
on the neutralino composition.
Requiring these two mechanisms to give rise to the full dark matter
energy density can be recast in a simple quadratic equation for $m_{3/2}$:
\begin{equation}
B\; m_{3/2}^2 - \Omega_{CDM} h^2 \;m_{3/2} + A \; T_{RH} = 0
\end{equation}
which has two real roots for $m_{3/2} $ only when
\begin{equation}
T_{RH} \leq \frac{(\Omega_{CDM} h^2)^2}{4 A B}
\label{eq:TRHupperboundanym12}
\end{equation}
and this value provides the maximal possible reheating temperature for any gravitino mass. 
Note that the r.h.s. of eq.~(\ref{eq:TRHupperboundanym12}) depends only on the
pMSSM parameters through the $A$ and $B$ coefficients.
The maximal reheating temperature is realized when the CDM density is provided equally by the two 
production mechanisms and the gravitino mass is fixed to the value
\begin{equation}
m_{3/2} = \frac{\Omega_{CDM} h^2}{2 B} = \frac{\Omega_{CDM} h^2}{\Omega_\chi h^2} \; \frac{M_\chi}{2}\; .
\label{eq:gravmassbound}
\end{equation}
If $ \Omega_\chi h^2 < \Omega_{CDM} h^2/2 $, this particular value of $ m_{3/2} $
is excluded by the requirement $ m_{3/2} < M_\chi $. For values of the gravitino mass lower than in
eq.~(\ref{eq:gravmassbound}), the bound on  $ T_{RH} $ gets stronger. Other cosmological constraints
restrict the region of large $m_{3/2} $ and in general in any point of the pMSSM parameter space set
a stronger bound on $T_{RH}$ compared to that given by eq.~(\ref{eq:TRHupperboundanym12}).
This constraint is given by
\begin{equation}
T_{RH} \leq \frac{m_{3/2}}{A} \left( \Omega_{CDM} h^2 - B m_{3/2} \right) \; ,
\label{eq:TRHbound}
\end{equation}
and it depends on both the gravitino mass and the pMSSM parameters.

The reheating temperature of the Universe is an important parameter in the cosmological evolution
and depends on both the particular inflationary model and the reheating process after inflation.
It strongly influences the possible mechanisms of baryogenesis, needed to generate the
present baryon asymmetry starting from a symmetric initial state. In particular, the thermal
leptogenesis mechanism~\cite{Fukugita:1986hr} relies on the presence of heavy right-handed Majorana
neutrinos in the thermal bath. These decay out-of-equilibrium giving rise to a non-vanishing lepton 
and then a baryon asymmetry.

Thermal leptogenesis can produce the observed baryon number only if the reheating temperature
is sufficiently high, above approximately $ 2 \times 10^9 $ GeV~\cite{Giudice:2003jh, Buchmuller:2004nz}.
With a mild tuning in the seesaw formula and exploiting flavor effects, this bound can be relaxed by an order of
magnitude~\cite{DiBari:2012fz, Fong:2013wr} and even more so in case of resonant CP violation, due to nearly
degenerate RH neutrino masses~\cite{Flanz:1996fb, Pilaftsis:1997jf}.
While these bounds are not intrinsic to the gravitino DM hypothesis,
they are a desirable addition to Big Bang cosmology. We will discuss how the requirement of high $ T_{RH} $ can
be used in these specific scenario to place strong constraints on the SUSY spectrum, when it is compared to the
reheating temperature needed to generate the correct gravitino DM abundance.  
Indeed, as can be inferred from eq.~(\ref{eq:thermalprod}), light degenerate gauginos reduce gravitino production
and allow for the highest possible $ T_{RH} $  around $10^9 \mbox{GeV} $~\cite{Olechowski:2009bd,Covi:2010au}.
This makes possible to interpret collider data as tests of the simplest thermal leptogenesis hypothesis.

\subsection{Cosmological constraints}
\label{sec:cosmo}

The requirement of a consistent cosmology sets important constraints on models with a
very weakly-interacting particle like the gravitino. Indeed, even if the gravitino is not the
LSP and therefore unstable, it may decay very late in the cosmological history leading to the
disruption of Big Bang Nucleosynthesis (BBN), the so-called
"gravitino problem"~\cite{Weinberg:1982zq,Khlopov:1984pf,Ellis:1984er}. It is interesting to
point out that these scenarios may be seen as an asset, since these late decays might solve
another outstanding problem, that of the amount of cosmic Lithium formed after the
Big Bang~\cite{Cyburt:2013fda}.
In our scenario the gravitino is the LSP and so the long-lived particle becomes the 
neutralino NLSP. Due to the Planck scale suppression, the NLSP lifetime can easily
exceed 1~s and the decay of the out-of-equilibrium NLSPs again takes place during BBN,
thus affecting the abundance of light elements~\cite{Jedamzik:2009uy}.
For a neutral long-lived particle like the neutralino, the main processes affecting BBN are 
photo- or hadro-dissociation, which depend on the number density of the decaying particle
and its branching fraction into photon or hadrons. For short lifetimes the constraints from
hadro-dissociation are stronger than those given by photo-dissociation, which takes over
at $\tau > 10^{4}-10^6 $ s \cite{Kawasaki:2004qu, Jedamzik:2006xz}. The constraints for neutral
relic with different values of the hadronic branching ratio and decaying particle mass
were computed by running a BBN code including the neutral particle decay channel into quark and
antiquark pairs in~\cite{Jedamzik:2006xz}. Here, we use those generic constraints and translate them
to the special case of the neutralino, similarly as what was done in \cite{Covi:2009bk,CahillRowley:2012cb}.
In order to do that we need to compute the decaying branching fractions and the freeze-out density
of the NLSP for every point of the pMSSM parameter space.

The decay rates of a general neutralino NLSP into a gravitino LSP were calculated analytically 
including  all possible three-body decays in~\cite{Covi:2009bk}. The neutralino decays
dominantly into a gravitino and a gauge boson. The hadronic channel into gravitino with a 
quark and antiquark pair arises most of the time from an intermediate gauge or Higgs boson, as long 
as the scalar quarks are not too light. So the most important hadronic channel for the neutralino
consists of a three-body decay, where part of the energy is carried away by the inert gravitino, instead
of the simple two-body decay into quark and antiquark, which was assumed in~\cite{Jedamzik:2006xz}
for the neutral relic. Therefore the quark-pair momentum distribution in neutralino decay is softer than 
in a two-body decay and we therefore regard the limits given in \cite{Jedamzik:2006xz} as possibly 
too strong for our scenario.
Indeed, if the intermediate gauge boson is on-shell before decaying into a quark-antiquark pair, the 
number of hadronic particles produced is expected to be independent of the gauge boson energy or 
momentum. Moreover, if the hadrons produced thermalise before interacting with light elements, any
dependence on the initial spectrum or the mass of the mother particle is washed out. In this spirit, the
work of ref.~\cite{CahillRowley:2012cb} uses the bounds of Jedamzik for a decaying particle mass of 100~GeV, 
disregarding the dependence on this mass and correcting the electromagnetic bounds by subtracting 
the energy carried away by the gravitino.
In this study we implement both sets of constraints on our scans: first the limits given by
\cite{Jedamzik:2006xz} as a function of the hadronic branching ratio and decaying particle mass,
and then the slightly weaker limits of ref.~\cite{CahillRowley:2012cb}.
Indeed, since in this analysis we include also regions of the parameter space where the $Z$ or Higgs
boson cannot be on shell, i.e. the region where $ M_1 - m_{3/2} < M_Z, M_h $,  we consider the
latter bounds as too conservative in some part of the parameter space.
The comparison between the two sets of bounds, with one appearing to be too strong and the other too 
weak in part of the parameter space, provides us with an estimation on the sensitivity of the results on
the precise implementation of these constraints.

In general, BBN bounds exclude regions where the density of the decaying particle as a function
of its lifetime is too large. Indeed in the early stages of BBN, corresponding to lifetimes below 10$^2$~s,
the constraints are very weak, or even non-existent, while after the production of light elements, like 
Deuterium, Helium or Lithium, the effect of hadro-dissociation becomes stronger. The lifetime of the 
NLSP in our scenario is always related to the gravitino mass, since for a light gravitino its couplings 
are proportional to $1/m_{3/2} $. These bounds can be translated in general into an upper 
bound on the gravitino mass, as a function of the NLSP density at freeze-out.
For a bino-like neutralino NLSP lifetime, above the $Z$ threshold, we have
\begin{equation}
\Gamma_{\tilde{\chi}_1^0 \rightarrow \gamma/Z \psi_{3/2}} =
\frac{1}{48\pi M_P^2} \frac{M_{\chi}^5}{m_{3/2}^2} =
(57 \mbox{s} )^{-1} \left(\frac{M_{\chi}}{1\;\mbox{TeV}}\right)^{5}   
 \left(\frac{m_{3/2}}{10\;\mbox{GeV}}\right)^{-2}\; .
 \label{eq:NLSPlifetime}
\end{equation}
For gravitino masses larger than about $1$~GeV, the NLSP typically decays during BBN and affects its
predictions. It is possible to shorten the NLSP lifetime by varying the value of the gravitino 
mass, independently of the decaying neutralino mass or composition.

\begin{figure}[t!]
\vspace*{-0.35cm}
\begin{center}
\includegraphics[width=0.90\columnwidth]{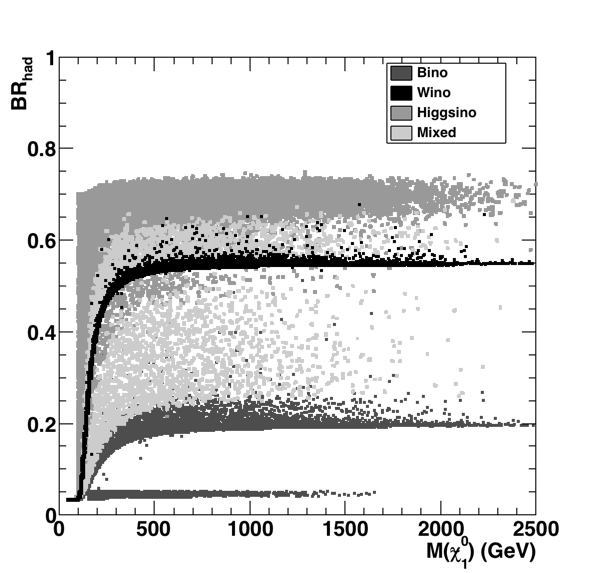}
\end{center}
\caption{Neutralino hadronic decay branching fraction as a function of the neutralino mass for the accepted 
pMSSM scan points, with the nature of the neutralino having the largest contribution in each region highlighted.}
\label{fig:MN1BRhad-BWH}
\end{figure}

We use the results of \cite{Covi:2009bk} to implement the hadronic constraints on the model, deriving 
the branching fraction for the neutralino decays into gravitino and quark-antiquark pair for each point
of the pMSSM parameter space scan. This branching fraction is shown as a function of the neutralino mass
for our pMSSM scan points in Figure~\ref{fig:MN1BRhad-BWH}. The nature of the neutralino is highlighted,
with the bino-, wino- and higgsino-like components defined as the points with the corresponding entry
squared in the neutralino mixing matrix exceeding 0.75. The minimal hadronic branching fraction for fixed 
neutralino mass, is obtained for the neutralino composition of a pure photino, since in this case the 
contributions through intermediate $Z$ and Higgses are absent.

We obtain the neutralino branching fraction into hadrons as follows.
First, we estimate the neutralino lifetime considering the decay into the gravitino and a photon, 
which is always open, keeping all kinematical factors in this channel. Then, we compute 
all the other decay widths, e.g. into $Z$ and $h$, by setting the gravitino mass to zero in the
kinematical factors. Since the photon or another open decay channel dominates, when thresholds 
are encountered, this approximation holds below the per-mille level in all points and gives a
rather conservative estimate of the hadronic branching fraction.
Next, we compute the decay rate into the hadronic channel taking into account the neutralino
composition and the presence of off-shell or on-shell intermediate bosons. If the intermediate
particles are on-shell, the result is well approximated by the decay rate into that particle
times the corresponding branching fraction. But our results extend also into the regions with
off-shell intermediate states. In this case, we integrate the branching fraction for energies of
the quark-antiquark pairs above their mass threshold. As discussed in \cite{Covi:2009bk} this
gives,  below the $Z$ threshold, a result three times larger than taking a threshold of 2~GeV,
corresponding to the production in the decay of hadrons instead of mesons.
At leading order the branching fraction is independent of the gravitino mass, since
all rates are proportional to the same factor $ 1/m_{3/2}^2 $. Only for large gravitino masses
near thresholds does this parameter play an important role also in the phase-space factors.

Photo-dissociation constraints are very similar in most of the parameter space and mostly independent
on the specific values of the purely electromagnetic branching ratios, since the decay into the
gravitino and a photon has always a branching fraction larger than 30\% and the hadronic bounds
in ref.~\cite{Jedamzik:2006xz} include the photo-dissociation effects from charged hadrons.

\begin{figure}[t!]
\vspace*{-0.35cm}
\begin{center}
\begin{tabular}{c}
\includegraphics[width=0.90\columnwidth]{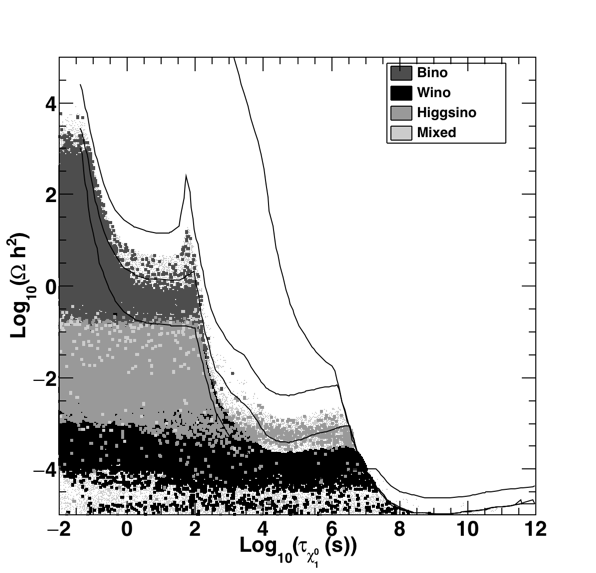} \\
\includegraphics[width=0.90\columnwidth]{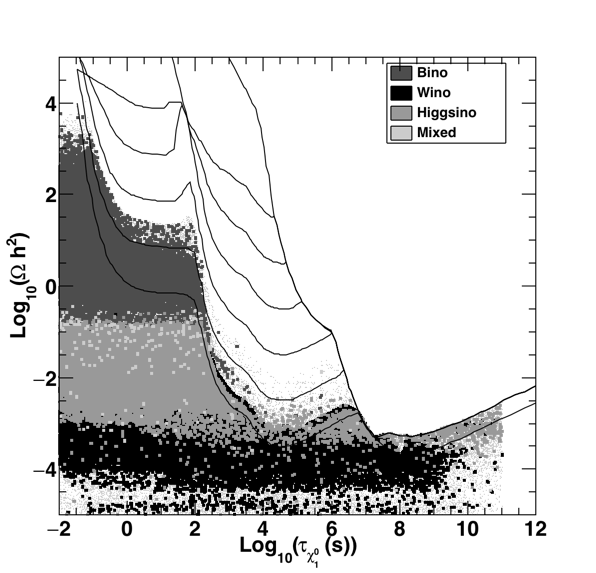} \\
\end{tabular}
\end{center}
\caption{Neutralino relic density as a function of the neutralino lifetime for the accepted pMSSM 
scan points, with the nature of the neutralino having the largest contribution in each region highlighted.
The BBN constraints obtained following ref.~\cite{Jedamzik:2006xz} (top) and ref.~\cite{CahillRowley:2012cb}
(bottom) are shown by the continuous lines corresponding to limits for neutralino hadronic branching 
fractions of 100\%, 10\%, 1\% and 0\% from bottom to top.}
\label{fig:TauN1Oh2-BWH}
\end{figure}

\begin{figure}[t!]
\vspace*{-0.35cm}
\begin{center}
\begin{tabular}{c}
\includegraphics[width=0.90\columnwidth]{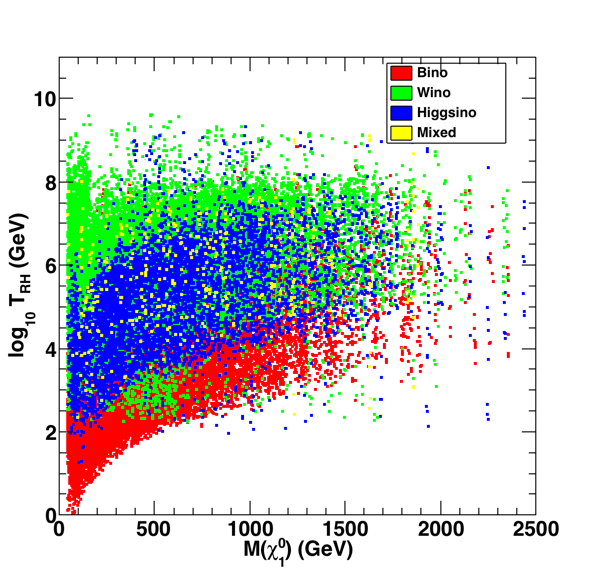} \\
\includegraphics[width=0.90\columnwidth]{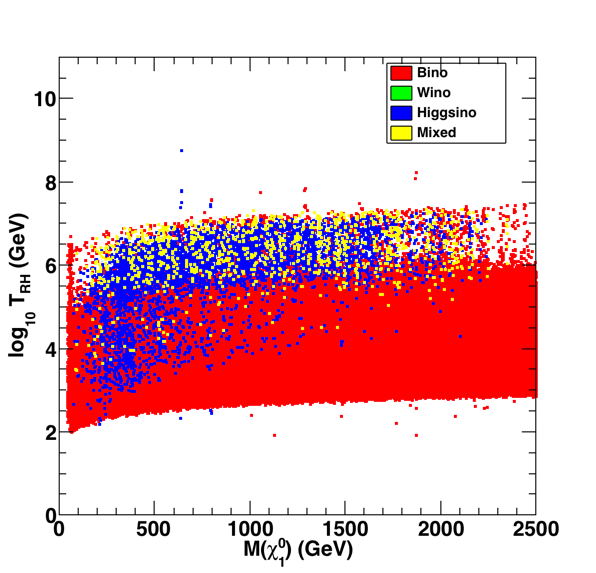} \\
\end{tabular}
\end{center}
\caption{Reheating temperature as a function of the NLSP neutralino mass for the accepted pMSSM (top) 
and cMSSM (bottom) scan points, with the nature of the neutralino having the largest contribution in each
region highlighted.}
\label{fig:MN1RHT-BWH}
\end{figure}

The calculation of the cosmological bounds require to implement the number density of the neutralino 
NLSP before decay. This is evaluated by considering the freeze-out process of the neutralino and
solving the corresponding Boltzmann equation using the {\tt SuperIso Relic} code~\cite{Arbey:2009gu}.
The neutralino NLSP density is determined by the 19 pMSSM parameters and it is independent
of the gravitino mass. Therefore we can find iteratively the largest NLSP lifetime and thus the
largest gravitino mass compatible with BBN and the NLSP number density for a set of pMSSM parameters,
as discussed in the next section.
The maximal allowed value for the gravitino mass at any chosen point in the pMSSM parameter space is
usually smaller than the general upper bound given in eq.~(\ref{eq:gravmassbound}) and therefore we
can apply eq.~(\ref{eq:TRHbound}) to obtain a tighter bound on the reheating temperature $T_{RH}$.
The distribution of the neutralino relic density as a function of its lifetime for the pMSSM scan
points is given is Figure~\ref{fig:TauN1Oh2-BWH}.
The dependence of the relic density of the neutralino from its composition clearly appears: a
bino neutralino has naturally a larger abundance and  therefore as an NLSP must have a shorter 
lifetime than a Wino and Higgsino neutralino in order to not disrupt BBN. We expect that
wino/higgsino-like NLSPs can allow for a larger maximal gravitino mass, and so also a larger reheating
temperature than a bino NLSP. This can be verified on our scans by studying the distribution of the
reheating temperature as a function of the NLSP mass and selecting the NLSP nature.
Figure~\ref{fig:MN1RHT-BWH} shows the results obtained constrasting the case of the pMSSM scan to those
of the cMSSM. In the pMSSM parameter space, points with wino- and higgsino-like NLSP accomodate values
of $T_{RH}$ in the range 10$^6$ - 5$\times$10$^9$~GeV. On the contrary, the bino-like population of the
cMSSM points does not offer solutions with $T_{RH}$ values in excess of 10$^7$~GeV.

For neutralino lifetimes longer than $10^{10}$~s, constraints from CMB distortion also play a role and
start to be more stringent than the BBN constraints, see e.g.~\cite{Hasenkamp:2013opa}. 
Since we can obtain such extremely large lifetime values only for very few fine-tuned points of our
parameter scan, as in the degenerate gravitino-NLSP scenario~\cite{Boubekeur:2010nt}, we implement an upper
cut at $\tau= 10^{11}$~s.

\section{Analysis}

\subsection{MSSM Scans}

The phenomenological MSSM (pMSSM) is the most general MSSM scenario with R-parity, CP
conservation and minimal flavour violation~\cite{Djouadi:1998di} and its properties are
defined by 19 independent parameters. This analysis is based on a flat scan where these parameters
are varied within the ranges given in Table~\ref{tab:paramSUSY}. The gravitino mass is added
as an additional parameter. The ranges of the SUSY particle masses have been made broader
compared to previous studies~\cite{Arbey:2011un,Arbey:2011aa,CahillRowley:2012cb} to achieve a
better assessment of the capability of the LHC at 13-14 TeV, which will probe the masses of
strongly interacting SUSY particles up to 2-3~TeV. For this analysis a total of 7$\times$10$^{7}$
pMSSM points have been generated and analysed.

\begin{table}
\begin{center}
\begin{tabular}{|c|c|}
\hline
~~~~Parameter~~~~ & ~~~~~~~~Range~~~~~~~~ \\
\hline\hline
$\tan\beta$ & [1, 60]\\
\hline
$M_A$ & [50, 5000] \\
\hline
$M_1$ & [-5000, 5000] \\
\hline
$M_2$ & [-5000, 5000] \\
\hline
$M_3$ & [0, 5000] \\
\hline
$A_d=A_s=A_b$ & [-15000, 15000] \\
\hline
$A_u=A_c=A_t$ & [-15000, 15000] \\
\hline
$A_e=A_\mu=A_\tau$ & [-15000, 15000] \\
\hline
$\mu$ & [-5000, 5000] \\
\hline
$M_{\tilde{e}_L}=M_{\tilde{\mu}_L}$ & [0, 5000] \\
\hline
$M_{\tilde{e}_R}=M_{\tilde{\mu}_R}$ & [0, 5000] \\
\hline
$M_{\tilde{\tau}_L}$ & [0, 5000] \\
\hline
$M_{\tilde{\tau}_R}$ & [0, 5000] \\
\hline
$M_{\tilde{q}_{1L}}=M_{\tilde{q}_{2L}}$ & [0, 5000] \\
\hline
$M_{\tilde{q}_{3L}}$ & [0, 5000] \\
\hline
$M_{\tilde{u}_R}=M_{\tilde{c}_R}$ & [0, 5000] \\
\hline
$M_{\tilde{t}_R}$ & [0, 5000] \\
\hline
$M_{\tilde{d}_R}=M_{\tilde{s}_R}$ & [0, 5000] \\
\hline
$M_{\tilde{b}_R}$ & [0, 5000] \\
\hline
\end{tabular}
\caption{Range of the pMSSM parameters adopted in the scans (in GeV when applicable).
\label{tab:paramSUSY}}
\end{center}
\end{table}

In our analysis, the SUSY mass spectra and couplings are generated from the input pMSSM parameters
with {\tt SOFTSUSY 3.3.3} \cite{Allanach:2001kg}. Only points having the lightest neutralino,
$\tilde{\chi}^0_1$, to be the next lightest SUSY particle to the gravitino and the lightest
Higgs mass in the range 122 - 128~GeV are accepted. Particle decay widths are calculated using
{\tt HDECAY 5} \cite{Djouadi:1997yw} and {\tt SDECAY} \cite{Muhlleitner:2003vg}.

In order to contrast the gravitino phenomenology in the parameter space of the pMSSM to that of
constrained MSSM models, we also perform scans of the constrained MSSM (CMSSM)~\cite{Kane:1993td,Ellis:2002rp}
where we vary the $M_0$, $M_{1/2}$, $\tan \beta$ and $A_0$ parameters in the ranges indicated in
Table~\ref{tab:paramCMSSM}.
\begin{table}
\begin{center}
\begin{tabular}{|c|c|}
\hline
~~~~Parameter~~~~ & ~~~~~~~~Range~~~~~~~~ \\
\hline\hline
$\tan\beta$ & [2, 68]\\
\hline
$M_0$ & [0, 10000] \\
\hline
$M_{1/2}$ & [0, 10000] \\
\hline
$A_{0}$ & [-10000, 10000] \\
\hline
\end{tabular}
\caption{Range of the cMSSM parameters adopted in the scans (in GeV when applicable).
Both signs of $\mu$ have been considered.
\label{tab:paramCMSSM}}
\end{center}
\end{table}

For each accepted point, we compute the flavour observables and the muon anomalous magnetic moment with
{\tt SuperIso v3.3} \cite{Mahmoudi:2007vz,Mahmoudi:2008tp}, and the neutralino relic density with
{\tt SuperIso Relic} \cite{Arbey:2009gu}, where the gravitino related calculations have also been
implemented. To allow for comparisons between the gravitino LSP and the neutralino LSP scenarios,
we compute the neutralino direct detection scattering cross sections using the
{\tt MicrOMEGAs} code~\cite{Belanger:2008sj}. 

The constraints from the LHC SUSY searches at 7+8 and their projection to 14 TeV are obtained through
the analysis of inclusive SUSY events generated by {\tt PYTHIA 8.150} \cite{Sjostrand:2007gs} with the
CTEQ6L1 parton distribution functions \cite{Pumplin:2002vw}. SUSY particle production accompanied by
a hard jet, relevant to monojet searches, is simulated using {\tt MadGraph 5} \cite{Alwall:2011uj}
followed by {\tt PYTHIA} for hadronisation, as described in \cite{Arbey:2013iza}. The physics objects
of the signal events are obtained with a parametric simulation of the LHC detector response performed
with {\tt Delphes 3.0} \cite{deFavereau:2013fsa}.

For each pMSSM point, we infer the maximal gravitino mass consistent with the BBN constraints and
leading to a relic density $\Omega_{\tilde{G}} h^2 \sim 0.11$, in agreement with the Planck CMB data \cite{Ade:2013zuv}.
First, we derive the maximal neutralino lifetime using the BBN limits given in section~\ref{sec:cosmo} from the
computed neutralino relic density and hadronic branching ratio. The corresponding gravitino mass is then
obtained through an iterative procedure from the maximal NLSP lifetime.
Once the maximal gravitino mass is obtained, we randomly scan over the gravitino mass values within a range comprised 
between 1/1000 and 10 times the maximal gravitino mass and keep ten of such gravitino mass values for each pMSSM point.
If the maximal gravitino mass cannot be obtained, we vary randomly the gravitino mass logarithmically between 0.1~MeV 
and the neutralino mass.
We  then compute the gravitino relic density from NLSP decay as given in eq.~(\ref{eq:outeqprod}) and check
if it is smaller than or equal to the DM relic density. Since the neutralino relic density is diluted by the ratio
of the gravitino to the neutralino mass and the BBN bounds are stronger for larger NLSP density, this is always the case.
To retrieve a relic density compatible with the cosmological observations, we compute by inverting 
 eq.~(\ref{eq:thermalprod})  the reheating temperature, 
$T_{RH}$, required to increase the gravitino abundance up to the value corresponding to the Planck observations.

In addition, we include a set of points for which the lightest neutralino is a photino, i.e. a
mixture of bino and wino with the dominant decay mode to a gravitino and a photon. These states can be obtained
at tree level for $M_1=M_2$ and $|M_1| \ll |\mu|$. However, the neutralino mass matrix receives higher order
corrections, which modify this condition. Therefore, we use a specific scan where we first impose $M_1=M_2$,
and then iteratively adjust the value of $M_2$ to retrieve the pure photino solution.

\subsection{Collider and low-energy constraints}

The analysis of the viable MSSM parameter space requires the implementation of the available constraints from 
different sectors.

\begin{table}[t!]
  \begin{center}
    \begin{tabular}{|c|c|}
      \hline
      \multirow{5}{*}{Flavour} & $2.63\times10^{-4}<\mbox{BR}(B\to X_s\gamma) < 4.23\times10^{-4}$ \cite{Amhis:2012bh}\\
      & $1.3\times10^{-9}<\mbox{BR}(B_s\to \mu^+\mu^-)_{\rm untag} < 4.5\times10^{-9}$ \cite{Aaij:2013aka,Chatrchyan:2013bka,bsmm-comb}\\
      & $0.33\times10^{-4}<\mbox{BR}(B_u\to \tau\nu) < 1.95\times10^{-4}$ \cite{Adachi:2012mm,Lees:2012ju}\\
      & $4.7\times10^{-2}<\mbox{BR}(D_s\to \tau\nu) < 6.1\times10^{-2}$ \cite{Amhis:2012bh,Akeroyd:2009tn}\\
      & $0.985 < R_{\mu23} < 1.013$ \cite{Antonelli:2008jg}\\
      \hline
      $(g-2)_\mu$ & $-2.4\times10^{-9}< \delta a_\mu < 5.0\times10^{-9}$ \cite{Brown:2001mga,Arbey:2011un} \\
      \hline
    \end{tabular}
  \caption{Summary of the constraints from flavour physics and muon anomalous magnetic moment.}
  \label{tab:flavour} 
  \end{center}
\end{table}

First the constraints from low energy observables are imposed to the generated points. Table~\ref{tab:flavour} summarises
the most recent values for the low energy 95\% C.L. limits, obtained by including experimental and theoretical uncertainties.
In particular, the branching fraction of the decay $B_s\to\mu^+\mu^-$ is sensitive to scalar and pseudoscalar operators, and
provides us with a bound on the $\tan\beta$ and $M_A$ parameters~\cite{Arbey:2012ax}. The value of the inclusive BR$(B\to X_s \gamma)$
is also sensitive to charged Higgs and charginos/stops contributions at loop level, thus constraining $\tan\beta$, $M_A$,
as well as the stop and chargino masses. These constraints remove $\simeq$~12\% of the accepted pMSSM points in our
scans.

With $\simeq$25~fb$^{-1}$ of statistics collected by both the ATLAS and CMS experiments at 7+8~TeV during the LHC Run-1, a vast
array of searches for the production and decay of new particles has been performed. In particular, the analysis of channels
with missing transverse energy (MET), including jets + MET, leptons + MET and monojets, have excluded a sizeable fraction of
the MSSM parameter space. With the start of Run-2 at 13~TeV and the perspectives of increased luminosity, the LHC will put SUSY
scenarios through a crucial, although possibly not yet definitive, test. 
\begin{table}
  \begin{center}
    \begin{tabular}{|l|c|c|c|}
      \hline
      Channel & Experiment & Sensitivity & Ref. \\
      \hline
      jets + MET           & ATLAS & $\tilde g$, $\tilde q$ & \cite{ATLAS-CONF-2013-047} \\
      2 $\ell$ + jets + MET & ATLAS & $\tilde t$, $\tilde q$ & \cite{Aad:2014qaa} \\
      1 $\ell$ + b-jet(s) + MET & ATLAS & $\tilde t$ & \cite{ATLAS-CONF-2013-037} \\
      $b$-jets + MET       & ATLAS & $\tilde t$, $\tilde b$ & \cite{ATLAS-CONF-2013-053} \\
      \hline
      2 $\ell$ + MET       & ATLAS & $\tilde \chi^0$ $\tilde \chi^{\pm}$ & \cite{ATLAS-CONF-2013-049} \\ 
      3 $\ell$ + MET       & ATLAS & $\tilde \chi^0$ $\tilde \chi^{\pm}$ & \cite{ATLAS-CONF-2013-035} \\
      1 $\ell$ + $bb$ + MET  & ATLAS & $\tilde \chi^0$ $\tilde \chi^{\pm}$ & \cite{ATLAS-CONF-2013-093} \\
      \hline
      monojet + MET        & ATLAS & $\tilde \chi \tilde \chi$, $\tilde q \tilde q$ & \cite{ATLAS-CONF-2013-068} \\  
      monojet + MET        & CMS   & $\tilde \chi \tilde \chi$, $\tilde q \tilde q$ & \cite{CMS-EXO-12-048} \\
      \hline
      $H$/$A \rightarrow \tau \tau$ & CMS & $H$, $A$ & \cite{CMS-HIG-13-021} \\
      \hline
    \end{tabular}
  \caption{Summary of the analyses used to assess the observability of the pMSSM points by the LHC SUSY searches.}
  \label{tab:lhc-susy}
  \end{center}
\end{table}
In this study, we consider the bounds obtained from the analyses summarised in Table~\ref{tab:lhc-susy}. Although these
are not exhaustive of the signal topologies investigated by the ATLAS and CMS searches, they cover the SUSY signatures
with the highest sensitivity and are largely uncorrelated. For each accepted pMSSM point, we simulate an inclusive SUSY
event sample and perform a parametric simulation for the event reconstruction, as discussed above. 
Signal selection cuts corresponding to each of the analyses are applied to these simulated signal events. The number of
SM background events in the signal regions are taken from the estimates reported by the experiments. The 95\% confidence
level (C.L.) exclusion in presence of background only is determined using the CLs method~\cite{Read:2002hq}. 

\begin{figure}[t!]
\begin{center}
\includegraphics[width=0.90\columnwidth]{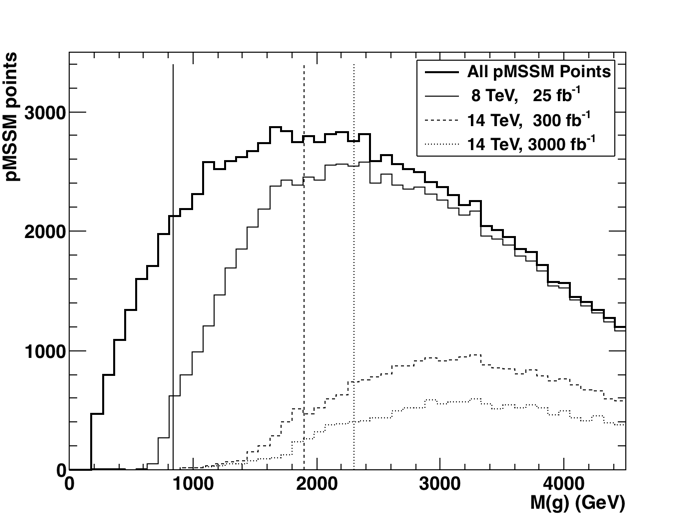}
\caption{Distribution of the gluino mass from the pMSSM scans with LSP
  gravitino and NLSP neutralino showing all the accepted points (black line)
  and those not excluded with 25~fb$^{-1}$ of 7+8~TeV data (grey line),
  300~fb$^{-1}$ (dashed line) and 3000~fb$^{-1}$ of 14~TeV data (dotted line).
  The vertical lines indicate the masses at which more than 95\% of our scan
  points with gluino mass below those values are, or will be, excluded by the
  LHC data.} 
\label{fig:MGL-LHC}
\end{center}
\end{figure}
These results are projected to 14~TeV for 300 and 3000~fb$^{-1}$ of integrated luminosity, by generating
events at 14~TeV and rescaling the 8~TeV backgrounds by the corresponding increase in cross section and signal
cut acceptance at the higher energy. The constraints derived from these searches exclude a significant fraction
of the gravitino pMSSM points, from 22\% at 8~TeV to 75\% and 85\% at 14~TeV with 300 and 3000~fb$^{-1}$, respectively.
The gluino mass is of particular importance in relation to the gravitino relic density and the reheating temperature.
Figure~\ref{fig:MGL-LHC} shows the distribution of the gluino mass for the points not excluded by the searches for the
different energies and data sets considered here. The gluino mass values at which more than 95\% of the scan points
below that mass are rejected by the LHC searches, or will be in case of a negative result, is $M_{\tilde{g}}$ = 840, 1900
and 2300~GeV for 25~fb$^{-1}$ at 8~TeV, 300~fb$^{-1}$ and 3000~fb$^{-1}$ at 14~TeV, respectively.
We notice here that, for the general supersymmetric spectra realised in the pMSSM, values of the gluino 
mass below 1~TeV are still viable, and that the next runs will be able to extend the sensitivity beyond 2~TeV.

Since the gravitino LSP implies different constraints from dark matter, it is interesting to contrast the parameter
space allowed in this scenario to that of the MSSM with neutralino LSP, after the LHC searches.
The main difference between the two models arises from the dark matter relic density constraint, which severely
restricts the MSSM parameter space with neutralino LSP and modifies the occurrence of neutralinos of different nature
and the relations of the masses of the other SUSY particles to that of the lightest neutralino. 
In the neutralino LSP scenario, bino-like neutralinos are relatively disfavoured by loose relic density constraint,
because in general they lead to too large relic density. But imposing the tight relic density constraint highlights the
bino-like scenario, while the wino- and higgsino-like scenarios become less attractive at the LHC since in these two cases
the correct $\Omega_{\chi} h^2$ is naturally reached only for neutralino masses well beyond 1~TeV, thus pushing the SUSY
spectrum to high masses. In the wino scenario more complex decay chains of strongly-interacting SUSY particles occur also
more frequently. In the bino scenario, it is necessary to have coannihilations, which means SUSY particles with masses
close to that of the neutralino LSP are required. This has important consequences for the SUSY detectability at LHC, since
compressed spectra or SUSY particles at high mass scales become favoured by the tight relic density bound in the neutralino
DM scenario. 
\begin{figure}[t!]
\begin{center}
\includegraphics[width=0.99\columnwidth]{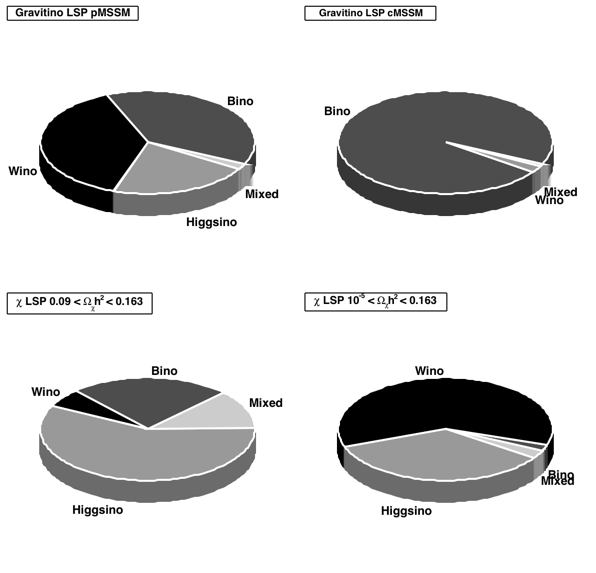}
\caption{Nature of the lightest neutralino for pMSSM (top left) and cMSSM (top right) points with gravitino LSP and
  pMSSM with neutralino LSP with tight (bottom left) and loose (bottom right) relic DM constraints.}
\label{fig:MSSM-N1}
\end{center}
\end{figure}

Here, we use two sets of $\Omega_{\chi} h^2$ constraints for the pMSSM points with neutralino LSP.
First, we apply a tight bound requiring the two values to be in agreement allowing for
systematic uncertainties, i.e.~$0.090 < \Omega_{\chi} h^2 < 0.163$, and assuming that the LSP
neutralino saturates the observed dark matter. We also consider a looser bound by simply requesting
that the neutralino relic density does not exceed the upper bound on the dark matter density from
the PLANCK CMB, i.e. $10^{-5} < \Omega_{\chi} h^2 < 0.163$,
again after accounting for systematic uncertainties. This allows for other sources of dark matter
in addition to the neutralino.
Of the pMSSM points we have studied fulfilling the flavour physics and low-energy data constraints,
1.7\% and 57\% satisfy the tight and loose constraints, respectively. It is therefore clear that some
amount of tuning is needed to obtain the right neutralino Dark Matter density.
Moreover, the tight $\Omega_{\chi} h^2$ constraint gives only 6.5\% of wino neutralinos, with 56\% higgsinos 
and 26\% bino. The loose constraint selects a sample of points with the lightest neutralino being
wino and higgsino in 61\% and 35\% of the cases, respectively. This should be contrasted with the
case of gravitino LSP models, characterised by an even distribution of points with different neutralino
nature, from bino (38\%) to wino (39\%) and higgsino (22\%) (see Figure~\ref{fig:MSSM-N1}). In contrast
to all these pMSSM scenarios, the gravitino model in the constrained MSSM is basically restricted to the
case of bino-like neutralinos, which acccounts for 97\% of the cMSSM accepted points, as shown in
Figure~\ref{fig:MSSM-N1}.

\section{Results}

The viable regions of the MSSM parameter space with gravitino LSP excluded by the LHC Run-1
searches and those projected for 300~fb$^{-1}$ and 3000~fb$^{-1}$ of data at 14~TeV, in the case no
excess of events above the SM backgrounds will be observed, have been studied for various combinations
of parameters. We present the results by computing the fractions of our scan points which are (will be)
excluded by the LHC searches given in Table~\ref{tab:lhc-susy}. Quantitatively, the values of
these fractions depend on the range of parameters used in the scans. However,
qualitatively the results hold when varying these ranges. The regions highlighted as being
deeply probed by the LHC searches remain so even if the SUSY particle masses are pushed
to higher values, although the fraction of points excluded will vary accordingly.
A first observation derived from these results is the existence of a significant parameter space with
neutralino NLSP when studying the pMSSM in sharp contrast with the results reported in previous studies
using the cMSSM and fixed values of $\tan \beta$ and $m_{\tilde{G}}$.
The sensitivity of the LHC data in falsifying the currently viable MSSM points with gravitino LSP can
be contrasted with that for models with neutralino LSP. In the previous section,
we have analysed the difference in the nature of the lightest neutralino for the two models.
\begin{figure}[t!]
\begin{center}
\begin{tabular}{c}
\includegraphics[width=0.75\columnwidth]{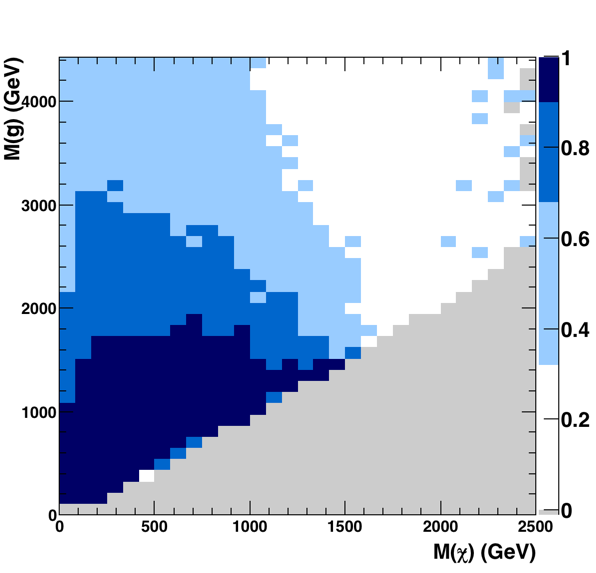} \\
\includegraphics[width=0.75\columnwidth]{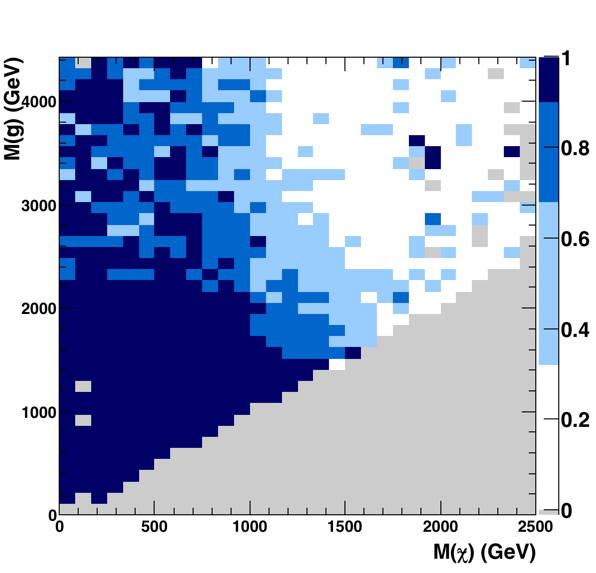} \\
\end{tabular}
\caption{Projected fractions of model points excluded by the LHC SUSY searches with
  300~fb$^{-1}$ at  14~TeV for the gravitino LSP (top) and the neutralino LSP scenario (bottom) 
  in the plane defined by the gluino and lightest neutralino masses. For neutralino LSP scenario
  only points consistent with the tight
  $\Omega_{\chi} h^2$ constraints and with neutralino scattering
  cross section $\sigma_{\chi p}^{SI}$
  consistent with the LUX data are considered. The regions in grey have no points found in our scans.}
\label{fig:MGLMN1-comp}
\end{center}
\end{figure}
Here, we concentrate on the fraction of points with SUSY masses up to 5~TeV excluded by the LHC searches.
Figure~\ref{fig:MGLMN1-comp} and Table~\ref{tab:MSSM-LHC} summarise our findings, with the regions in grey being
not accessible with the statistics of our scans. The distributions
of the viable points from the scans and their fraction excluded by a negative result with 300~fb$^{-1}$
at 14~TeV in the gravitino LSP and neutralino LSP scenarios with tight dark matter relic density
constraint are shown by the figure in the plane of the gluino mass versus the neutralino mass. 
We see here that the reach in neutralino mass is similar in the two cases and reaches approximately
1 TeV, while in the direction of the gluino mass, the LHC will be able to test better the large mass regions
in the case of a neutralino LSP and DM. We trace this back to the fact that in case of a light neutralino DM,
more supersymmetric particles have to be relatively light (e.g. the charged higgsino/winos for the
case of higgsino/wino neutralino) and can give rise to detectable signals even if the coloured states
are very heavy.

\begin{table}
 \begin{center}
    \begin{tabular}{|l|c|c|c|}
      \hline
      & $\tilde{G}$ LSP  & $\tilde{\chi}^0$ LSP & $\tilde{\chi}^0$ LSP \\
      &                & $.09< \Omega_{\chi}h^2<.163$ & $10^{-5}< \Omega_{\chi}h^2<.163$ \\
      \hline
      8 TeV 25 fb$^{-1}$  &                &                &              \\ 
      All                 &   0.218        &    0.100       &  0.188        \\ 
      Bino                &   0.273        &    0.268       &  0.312        \\ 
      Wino                &   0.144        &    0           &  0.145            \\ 
      Higgsino            &   0.245        &    0           &  0.249            \\ 
      \hline
      14 TeV 300 fb$^{-1}$  &               &                &              \\ 
      All                 &   0.745        &    0.533       &  0.694        \\ 
      Bino                &   0.835        &    0.851       &  0.864        \\ 
      Wino                &   0.614        &    0.035       &  0.615        \\ 
      Higgsino            &   0.808        &    0.343       &  0.811        \\ 
      \hline
      14 TeV 3 ab$^{-1}$   &                &                &               \\ 
      All                 &   0.845        &    0.745       &  0.806        \\ 
      Bino                &   0.917        &    0.956       &  0.927        \\ 
      Wino                &   0.733        &    0.212       &  0.736        \\ 
      Higgsino            &   0.902        &    0.631       &  0.907        \\ 
      \hline  
    \end{tabular}
  \caption{Fraction of pMSSM points with gravitino LSP and neutralino LSP fulfilling the
    flavour physics and low energy data constraints excluded by LHC searches.}
   \label{tab:MSSM-LHC}
  \end{center}
\end{table}

The fractions of points excluded by the analyses at 8, 14~TeV LHC and HL-LHC are given in Table~\ref{tab:MSSM-LHC} for
gravitino LSP and neutralino LSP with the tight and loose dark matter relic density constraints.
Since the composition of the sample of accepted points has significant differences in the
nature of the lightest neutralino, as discussed above, these fractions are also given restricting the
analysis to the points which have bino-, wino- and higgsino-like neutralino.
We observe that the fraction of excluded gravitino LSP points is significantly larger (by $\sim$22\% for the
Run-1 data and almost 75\% at 14~TeV) than that of models with neutralino
LSP and tight relic dark matter constraints. This is largely due to the effect of the higgsino-like
neutralino component in the neutralino LSP sample. The higgsino/wino relic density naturally reaches the
CMB value for neutralino LSP masses in the range 1.5-2.5~TeV. This pushes the SUSY spectrum to masses of
$\sim$2~TeV or more, i.e.\ at the limit of the LHC sensitivity or beyond. Therefore, if the $\Omega_{\chi} h^2$
constraints are enforced, the LHC sensitivity to higgsino- and wino-like neutralinos LSP scenarios collapses.
Indeed no points are excluded in this case from Run-1 data, as seen in Table~\ref{tab:MSSM-LHC}, and also
the future runs will not be able to cover the whole parameter space. 
Of course in case of the loose DM constraint, the neutralino mass is not pushed to such high values and
the LHC is more effective in testing the model, at a level similar to that observed for the gravitino LSP scenario. 
The LHC searches are also particularly sensitive to the photino points, as well as to the other bino-like solutions. 
The fractions of these points removed by the Run-1 data and the projections to 300 and 3000~fb$^{-1}$ of data 
at 14~TeV are 37, 92 and 98\%, respectively. 
In the case of the gravitino scenario in the cMSSM, the projection for 300~fb$^{-1}$ at 14~TeV has $\sim$34\% of the
points excluded if no signal is observed, when restricting the analysis to points with SUSY particle masses below
5~TeV to cover a parameter space comparable to that used for the pMSSM. The difference in the fraction of excluded
points is related to the distribution of the SUSY masses after applying all our selection criteria.

\begin{figure}[t!]
\begin{center}
\begin{tabular}{c}
\includegraphics[width=0.75\columnwidth]{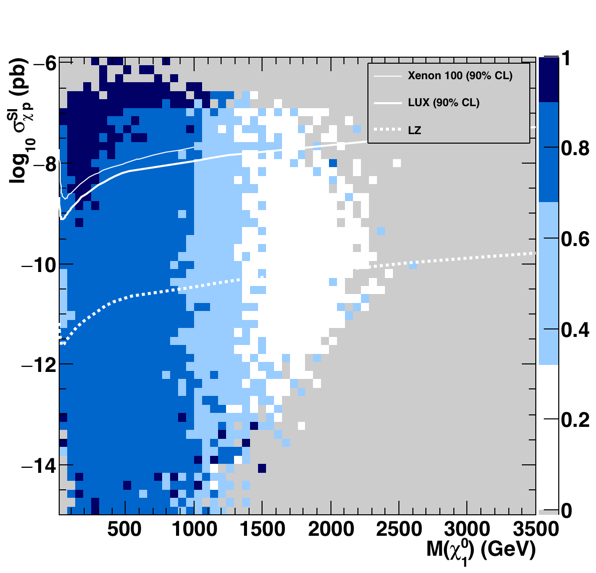} \\
\includegraphics[width=0.75\columnwidth]{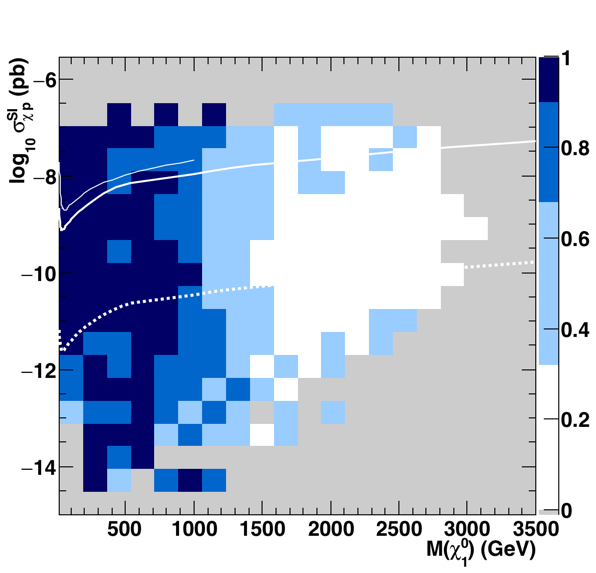} \\
\end{tabular}
\caption{Fractions of gravitino (top) and neutralino (bottom) LSP model points excluded by the LHC SUSY searches
  for the projection for 14~TeV  with 300~fb$^{-1}$ in the plane defined by the spin-independent neutralino
  scattering cross section and the neutralino mass. For the neutralino LSP the tight dark matter constraint is
  applied.}
\label{fig:MN1DD-LHC}
\end{center}
\end{figure}
The gravitino and neutralino LSP MSSM scenarios can be probed and differentiated by combining the LHC and
DM direct detection searches. In fact, in the case of gravitino LSP no signal from WIMP scattering on nucleons
is expected in the underground DM searches. On the contrary, if the neutralino LSP is the WIMP DM particle,
a large fraction of the points to which ATLAS and CMS will become sensitive at the LHC forthcoming runs are
also expected to give a signal at the LUX~\cite{Akerib:2013tjd} or LZ DM
experiments~\cite{Malling:2011va,Cushman:2013zza}. 
The distribution of the fraction of points excluded by LHC with 300~fb$^{-1}$ of 14~TeV data in the neutralino
scattering cross section vs.\ neutralino mass plane is shown in Figure~\ref{fig:MN1DD-LHC} for the
gravitino and neutralino LSP and DM scenarios.  We see here that, in the gravitino DM scenario with
an NLSP mass below 1 TeV,  a large part of the parameter points tested at LHC will also be tested
by LZ and in that case the presence of a positive signal at collider without a direct detection counterpart
will be a strong indication that the neutralino is not the LSP. For the case of a neutralino LSP and DM
instead, the direct detection experiments are able to test also regions outside of the LHC sensitivity
and detect heavy neutralinos, well beyond 1 TeV. Therefore, we observe that only a small fraction of points in
our scan are able to evade both the LHC and the direct detection constraints.
In quantitative terms, we find that 64 (62)\% of the gravitino LSP and 82 (80)\% of the neutralino LSP pMSSM 
points to which the LHC experiments will be sensitive after integrating 300~fb$^{-1}$ (3000~fb$^{-1}$) of data 
at 14~TeV could be also tested by the planned LZ
direct detection experiment, when the loose $\Omega_{\chi} h^2$ constraints are applied. In this case,
it may be appropriate to rescale the $\chi$ scattering cross section for the neutralino LSP case by the
ratio of the neutralino relic density to the CMB value. This lowers the fraction of points given above
to 42 and 39~\%, for 300 and 3000~fb$^{-1}$ respectively, in line with the result that we obtain when
enforcing the tight relic density constraints.

\begin{figure}[t!]
\begin{center}
\begin{tabular}{c}
\includegraphics[width=0.75\columnwidth]{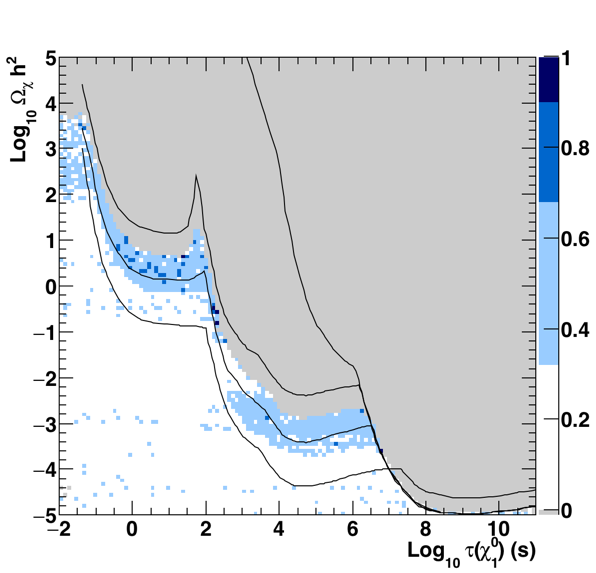} \\
\includegraphics[width=0.75\columnwidth]{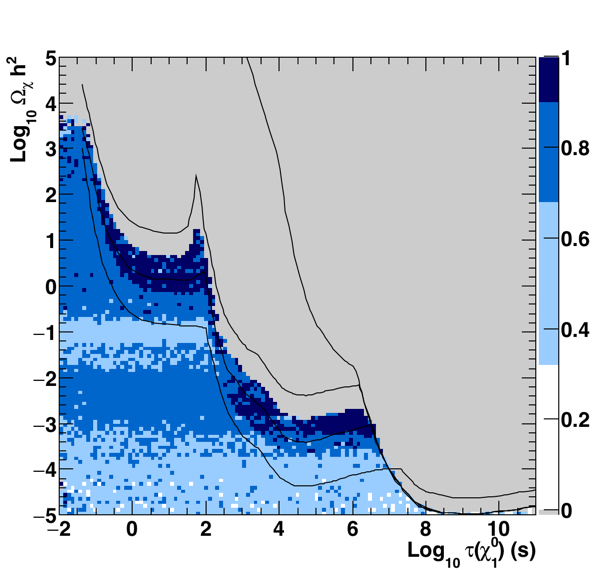} \\
\includegraphics[width=0.75\columnwidth]{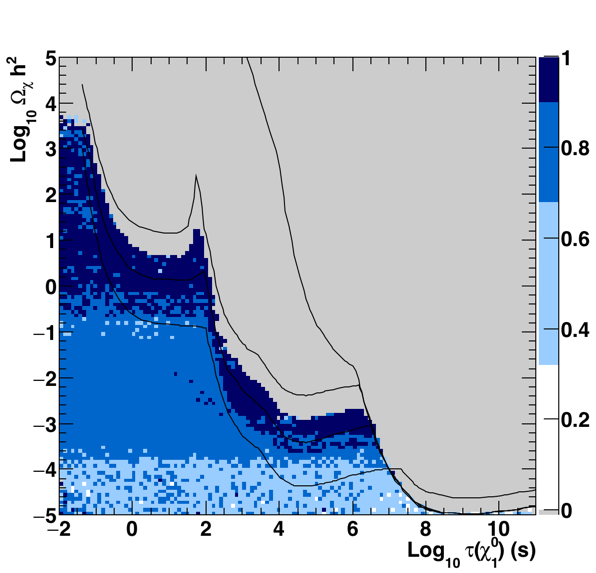} \\
\end{tabular}
\caption{Fractions of gravitino LSP model points excluded by the LHC SUSY searches with
  the 7+8~TeV data (top) and the projection for 14~TeV  with 300~fb$^{-1}$ (centre) and
  3000~fb$^{-1}$ (bottom) in the plane defined by the neutralino relic density and the
  neutralino lifetime. The BBN constraints obtained following ref.~\cite{Jedamzik:2006xz}
  are shown by the continuous lines corresponding to limits for neutralino hadronic branching 
  fractions of 100\%, 10\%, 1\% and 0\% from bottom to top.}
\label{fig:TauN1Oh2-LHC}
\end{center}
\end{figure}
\begin{figure}[t!]
\begin{center}
\begin{tabular}{c}
\includegraphics[width=0.75\columnwidth]{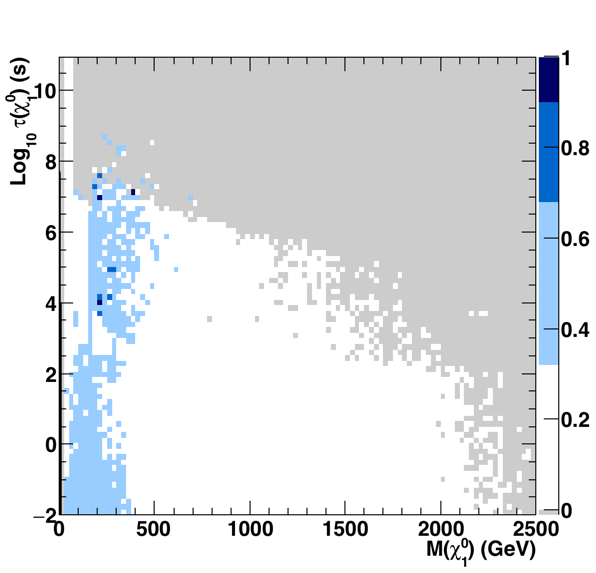} \\
\includegraphics[width=0.75\columnwidth]{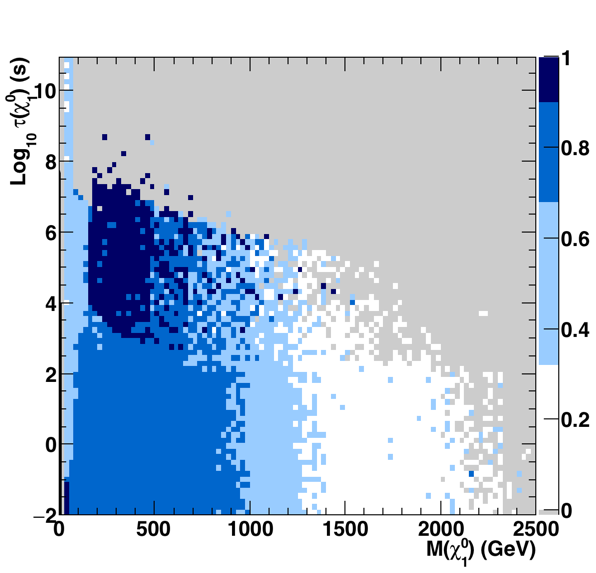} \\
\includegraphics[width=0.75\columnwidth]{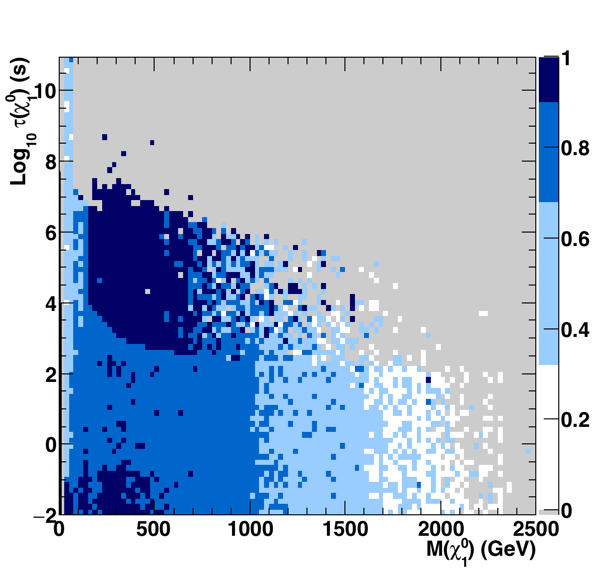} \\
\end{tabular}
\caption{Fractions of gravitino LSP model points excluded by the LHC SUSY searches with
  the 7+8~TeV data (top) and the projection for 14~TeV  with 300~fb$^{-1}$ (centre) and
  3000~fb$^{-1}$ (bottom) in the plane defined by the neutralino mass and the neutralino
  lifetime.}
\label{fig:MN1TauN1-LHC}
\end{center}
\end{figure}

Turning to the sensitivity over specific regions of the parameter space, the first set of variables
we analyse is the relation of the neutralino relic density, mass and lifetime
(see Figures~\ref{fig:TauN1Oh2-LHC} and \ref{fig:MN1TauN1-LHC}). The sensitivity to the neutralino
nature and mass is mapped, in part, onto these distributions. For example, the regions where the lightest 
neutralino is bino-like are preferentially excluded resulting in the largest sensitivity from the LHC Run-1 
data and the projections for the 14~TeV data being obtained for the parameter regions closest to the 
cosmological constraints in Figure~\ref{fig:TauN1Oh2-LHC}. 
Note on the other hand, comparing with Figure~\ref{fig:TauN1Oh2-BWH}, that also the region of 
wino NLSP will be well-tested in the future runs.
Moreover, the points with larger lifetime, just at the boundary of the constraint, correspond on average also 
to smaller neutralino mass and are more easily tested at a collider. Such a trend is clearly visible in  
Fig.~\ref{fig:MN1TauN1-LHC}.

In order to evaluate the sensitivity to the different assumptions used to derive the BBN constraints,
we compare the results obtained using the constraints of ref.~\cite{Jedamzik:2006xz}, which are used 
throughout our study, to those of ref.~\cite{CahillRowley:2012cb}. 
We find that the differences are minimal and restricted only to very particular points at the boundary.
So the fraction of pMSSM points with gravitino LSP fulfilling the flavour physics and low energy data 
constraints excluded by LHC searches for these two implementations of the BBN bounds agree to better 
than 5\%. The two samples of points differ mostly in the particular value of the maximal 
gravitino mass, which does not affect the LHC phenomenology. Moreover, since we are not scanning 
systematically on that mass, but taking a comparable number of points at the boundary, such a difference 
does not influence substantially the fraction of excluded points. 


\begin{figure}[t!]
\begin{center}
\begin{tabular}{c}
\includegraphics[width=0.75\columnwidth]{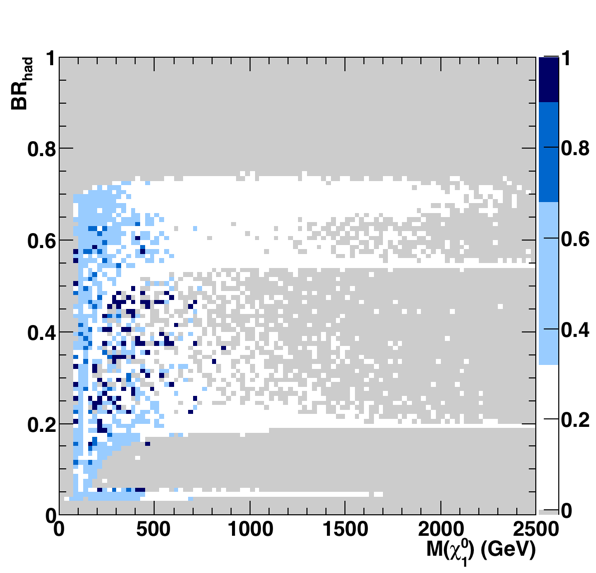} \\
\includegraphics[width=0.75\columnwidth]{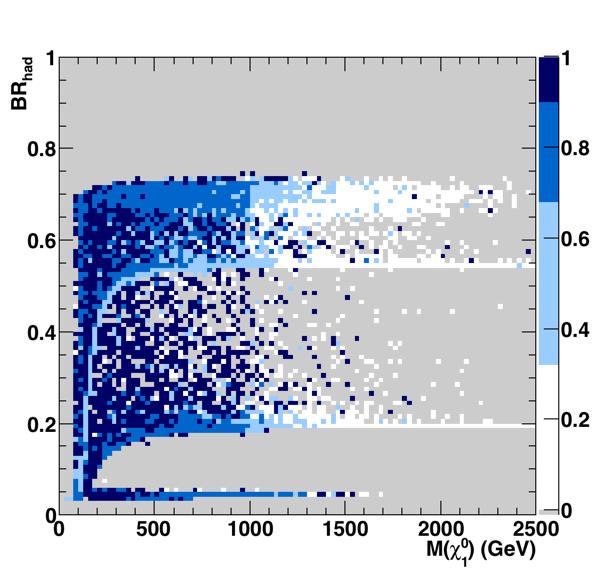} \\
\includegraphics[width=0.75\columnwidth]{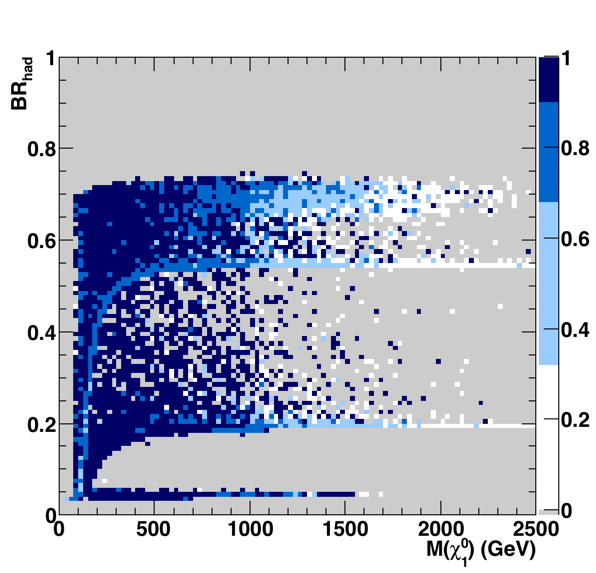} \\
\end{tabular}
\caption{Fractions of gravitino LSP points excluded by the LHC SUSY searches with the 7+8~TeV data (top) 
and the projection for 14~TeV  with 300~fb$^{-1}$ (centre) and 3000~fb$^{-1}$ (bottom) in the plane 
defined by the neutralino hadronic decay branching fraction and the neutralino mass. Note the photino 
solutions at small branching fraction values being almost completely excluded in case of no SUSY signal
at the HL-LHC.}
\label{fig:MN1BRHad-LHC}
\end{center}
\end{figure}

The fractions of excluded points in the plane of the hadronic branching fraction as a function of the neutralino mass
are shown in Figure~\ref{fig:MN1BRHad-LHC}. Here, it is interesting to observe that the photino solutions
can be largely excluded in the case the HL-LHC will not observe an excess of events in the SUSY searches:
97\% of the points in our scan are excluded with 3000~fb$^{-1}$ at 14~TeV, as shown in the right panel of
Figure~\ref{fig:MN1BRHad-LHC}. But in general the sensitivity of the LHC in the gravitino DM scenario is again 
mostly determined by the neutralino NLSP mass scale rather than by its composition.

\begin{figure}[t!]
\begin{center}
\begin{tabular}{c}
\includegraphics[width=0.75\columnwidth]{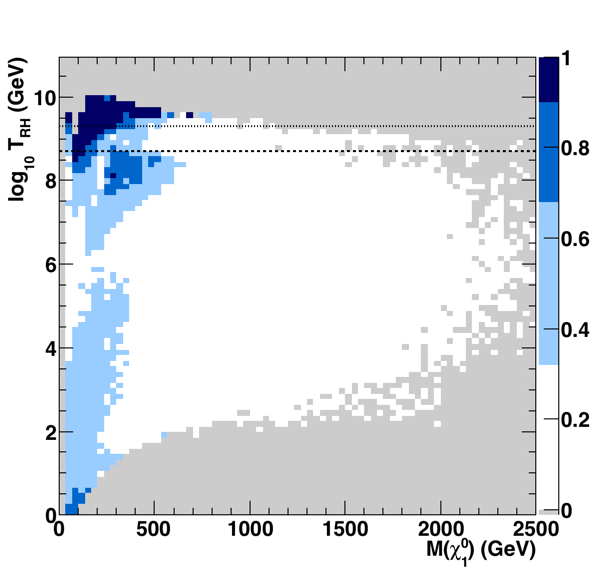} \\
\includegraphics[width=0.75\columnwidth]{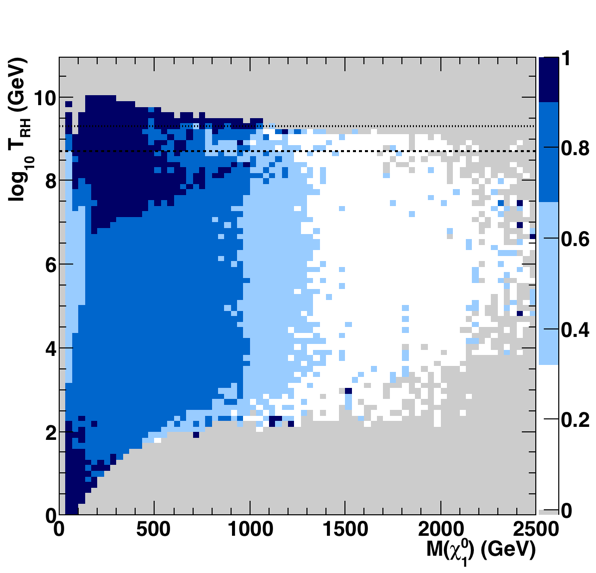} \\
\includegraphics[width=0.75\columnwidth]{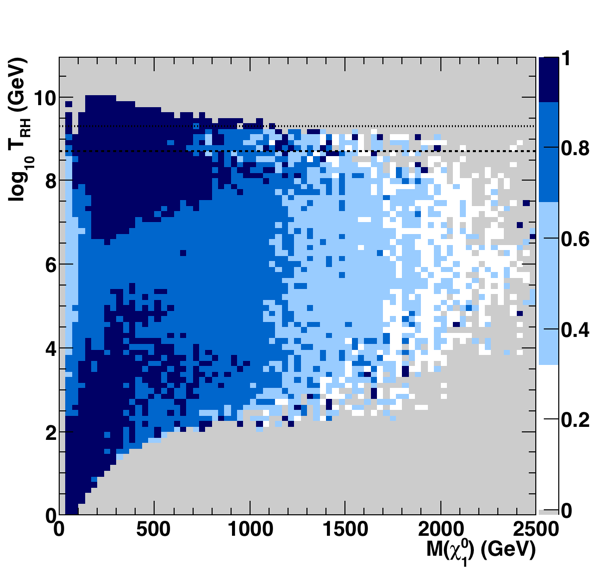} \\
\end{tabular}
\caption{Fractions of gravitino LSP points excluded by the LHC SUSY searches with the 7+8~TeV data (top) 
and the projection for 14~TeV  with 300~fb$^{-1}$ (centre) and 3000~fb$^{-1}$ (bottom) in the plane 
defined by the reheating temperature and the neutralino mass. The horizontal lines show the reheating 
temperature constraints from leptogenesis.}
\label{fig:MN1RHT-LHC}
\end{center}
\end{figure}

The LHC is also particularly sensitive to the test of gravitino LSP scenarios with relatively light NLSP
neutralinos and large reheating temperature, $T_{RH}$, close to the constraints derived from
thermal leptogenesis, for which we assume the lower bounds of $2 \times 10^9$~GeV~\cite{Giudice:2003jh, Buchmuller:2004nz}
or $5 \times 10^8$~GeV, in presence of flavour effects~\cite{DiBari:2012fz, Fong:2013wr}, as shown in
Figure~\ref{fig:MN1RHT-LHC}. Compared to cMSSM studies, which have essentially only bino $\chi$ NLSP as we have discussed above,
the pMSSM scans generated for this analysis have points with wino $\chi$ LSP that enable to increase the values of $T_{RH}$
reachable in this scenario (see Figure~\ref{fig:MN1RHT-BWH}) and make the LHC sensitivity compelling for their test.
These results are consistent with those obtained by the analysis of Ref.\cite{Covi:2009bk} that found a reheating
temperature of order of 10$^9$~GeV for the case of 1.25~TeV gluino and maximal gravitino mass around 70~GeV.
In our scan we also reach larger gravitino masses, i.e. also larger $T_{RH}$, since the neutralino relic density in
the pMSSM reaches lower values than those obtained in~\cite{Covi:2009bk}. 

Indeed a high reheating temperature is necessary to obtain the right gravitino abundance when the neutralino is light 
and has too low relic density to produce a sizeable gravitino population in its decay.
We see clearly in the figure also the complementary region of parameter space, where the whole gravitino density
is indeed generated by the light neutralino decay and therefore the reheating temperature has to be very low.
Both these two parameter regions will be completely tested in the next run of the LHC for neutralino masses
below 500-700~GeV.

\begin{figure}[t!]
\begin{center}
\begin{tabular}{c}
\includegraphics[width=0.75\columnwidth]{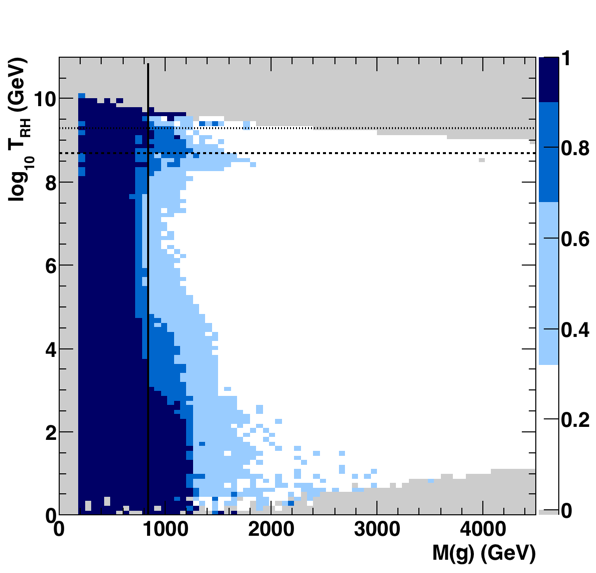} \\
\includegraphics[width=0.75\columnwidth]{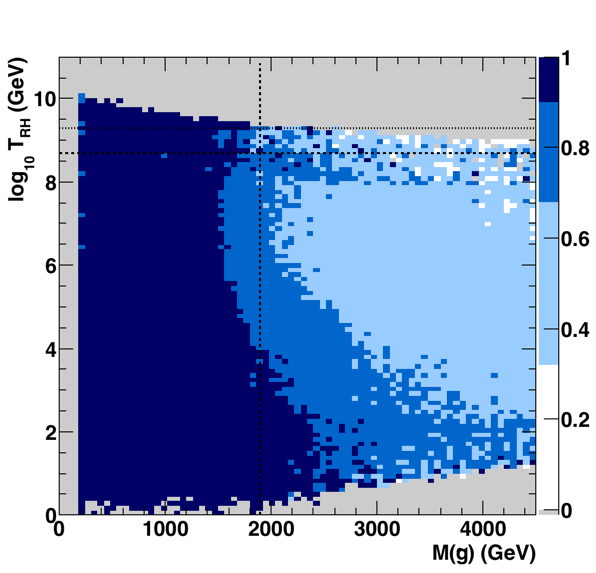} \\
\includegraphics[width=0.75\columnwidth]{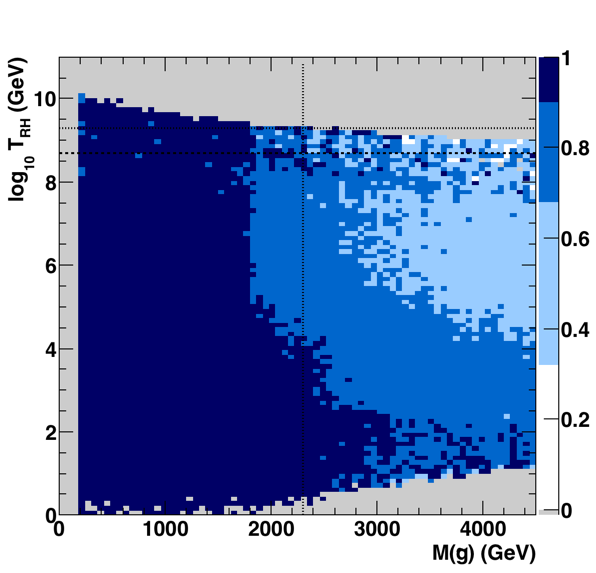} \\
\end{tabular}
\caption{Fractions of gravitino LSP points excluded by the LHC SUSY searches with the 7+8~TeV data (top) 
and the projection for 14~TeV  with 300~fb$^{-1}$ (centre) and 3000~fb$^{-1}$ (bottom) in the plane 
defined by the reheating temperature and the gluino mass. The horizontal lines show the reheating 
temperature constraints from leptogenesis and the vertical lines give the gluino mass value for which
more than 95\% of the points below it are excluded by the LHC SUSY searches.}
\label{fig:MGLRHT-LHC}
\end{center}
\end{figure}

\begin{figure}[t!]
\begin{center}
\begin{tabular}{c}
\includegraphics[width=0.75\columnwidth]{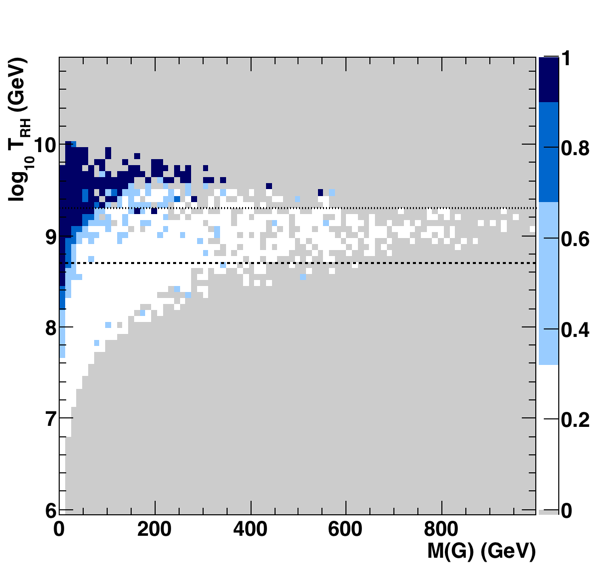} \\
\includegraphics[width=0.75\columnwidth]{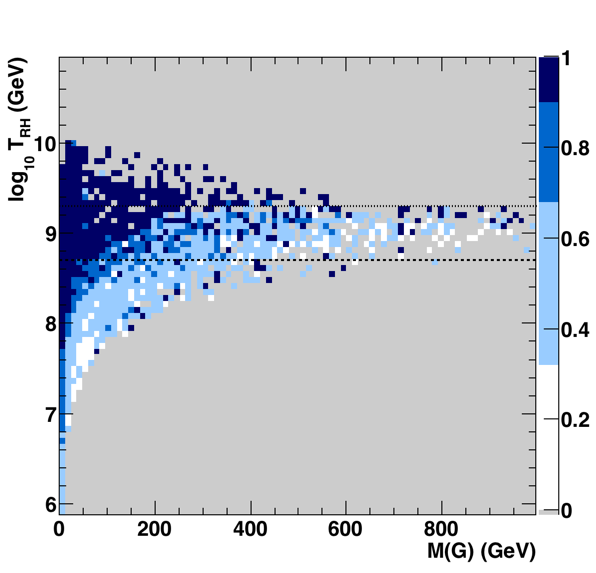} \\
\includegraphics[width=0.75\columnwidth]{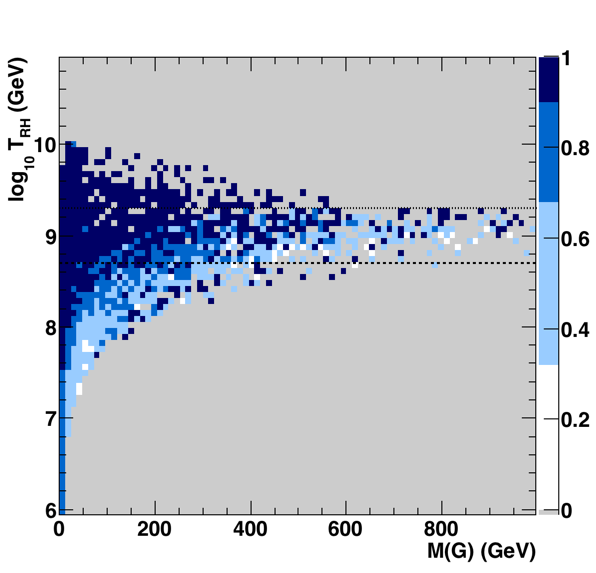} \\
\end{tabular}
\caption{Fractions of gravitino LSP points excluded by the LHC SUSY searches with the 7+8~TeV data (top) 
and the projection for 14~TeV  with 300~fb$^{-1}$ (centre) and 3000~fb$^{-1}$ (bottom) in the plane 
defined by the reheating temperature and the gravitino mass. The horizontal lines show the reheating 
temperature constraints from leptogenesis.}
\label{fig:MGRRHT-LHC}
\end{center}
\end{figure}

In these scenarios, the LHC sensitivity mostly comes from the constraints on the gluinos which have important
implications on the viable values of the reheating temperature. The interplay of the gravitino and gluino
masses and $T_{RH}$ for the points of our scan after imposing the current and expected LHC constraints in
absence of a signal is highlighted in Figures~\ref{fig:MGLRHT-LHC} and \ref{fig:MGRRHT-LHC}. 

From Fig.~\ref{fig:MGLRHT-LHC}, we can see clearly that the requirement of a specific reheating temperature
and the LHC exclusion on the gluino mass are completely orthogonal constraints. In the upper $T_{RH} $ region
we point out that the boundary of the sample points is determined by the overclosure bound $\Omega_{\tilde{G}} h^2 < 0.163 $:
since the gravitino density is increased both by large reheating temperature and gluino mass, as given in 
eq.~(\ref{eq:thermalprod}), such constraint imposes an anticorrelation between the maximal temperature allowed 
and the gluino mass.
Note that the fact that the low reheating temperature region is tested also at very large gluino masses, well beyond the 
LHC reach, is connected to the presence of a parameter region where the gravitinos are produced via neutralino decay 
and in that case neither the LHC constraints nor the gravitino density are derived from the coloured sector of the model.
In Fig.~\ref{fig:MGRRHT-LHC} we can see instead the points in our scan along the gravitino mass. Since we took 
mostly values of the gravitino mass around the BBN bounds, the distribution is not uniform in this direction.
We note that here the points corresponding to low $ T_{RH} $ have mostly a very small gravitino mass.
The particular behaviour of the upper bound is not due, in this case, directly to the overclosure bound, 
which would give $ T_{RH}^{max} \propto M_{\tilde{G}} $, but to the interplay of the overclosure bound with the 
BBN constraints. 

\begin{figure}[t!]
\begin{center}
\includegraphics[width=0.99\columnwidth]{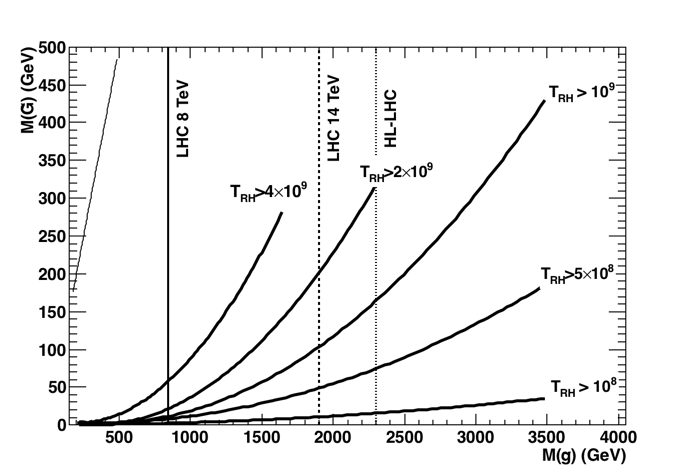}
\caption{Minimum values of the gravitino mass obtained in the gravitino LSP scans as a function of the gluino mass for
  different choices of the reheating temperature (in GeV). The values of $2 \times 10^{9}$ and $5 \times 10^8$~GeV
  correspond to the limits imposed by leptogenesis \cite{Giudice:2003jh, Buchmuller:2004nz, DiBari:2012fz, Fong:2013wr}.
  The thin diagonal line indicates the lower limit of the region where the gravitino is not the LSP. The vertical lines give
  the values of the gluino masses for which more than 95\% of the scan points below them are excluded by the LHC SUSY searches
  at 7+8~TeV (continuous) and projected at 14~TeV for 300 (dashed) and 3000~fb$^{-1}$ (dotted), if no signal is observed.}
\label{fig:MGLMGR-line}
\end{center}
\end{figure}

As we have discussed, the contribution of heavy gluinos to the production of gravitinos after inflation affects the
$\Omega_{\tilde{G}} h^2$ value, as can be seen from eq.~(\ref{eq:thermalprod}) and LHC bounds, if combined with the additional
requirement of successful thermal leptogenesis  may effectively test the resulting gravitino LSP
scenarios~\cite{Fujii:2003nr, Roszkowski:2004jd}. 
This is highlighted in Figure~\ref{fig:MGLMGR-line}, which summarises the results of our scans by giving the contours
for the minimum values of the gravitino masses obtained as a function of the gluino mass for different choices of $T_{RH}$.
This plot can be read in two complementary ways. First, by requiring a particular minimal reheating temperature together with
the relic density constraint, we can set strong bounds on the gluino and gravitino masses, since these determine the 
gravitino abundance in most of our parameter space. For gluinos heavier than 800~GeV, satisfying the present
LHC exclusion bounds, the gravitino mass has also to be substantial for reheating temperature above $10^9$~GeV.
Conversely, for each gravitino mass value, a choice of $ T_{RH} $  corresponds to an upper bound on the gluino
mass to avoid overclosure.
Then, by comparing these contours with the reheating temperature needed by thermal leptogenesis discussed above, 
we observe that the LHC data should either observe a gluino signal or exclude the gravitino DM scenarios with gravitino 
mass below 200-300~GeV and classical thermal leptogenesis, after accumulating sufficient statistics.
In this study the availability of the data set anticipated from the HL-LHC operation is of crucial importance, since the
sensitivity afforded by 3~ab$^{-1}$ of data at 14~TeV would cover the region compatible with the leptogenesis data up to at
least $M_{\tilde{G}} \simeq$ 300~GeV for a reheating temperature bound of $T_{RH} > 2 \times 10^9$~GeV and more for higher values
of $T_{RH}$. In this case, the lack of observation of a gluino in the LHC data would basically exclude most of the gravitino LSP
scenarios discussed here for the reheating temperatures required by thermal leptogenesis. 
Therefore, gravitino LSP models, in conjunction
with constraints derived from leptogenesis, offer a strong motivation for the HL-LHC program. Scenarios of flavoured
leptogenesis, allowing for lower $ T_{RH} \geq 5 \times 10^8$~GeV, may still accommodate gluino masses even larger than
those which can be probed at the HL-LHC, but could still be probed for gravitino masses below $\simeq$~60~GeV.

\section{Conclusions}

The search for dark matter is tightly connected to that of new physics beyond the SM. 
These searches bring together the efforts and the data accumulated by the LHC experiments but also the DM underground
direct searches. Gravitino Dark Matter represents a compelling scenario in Supersymmetry.
In this paper we have presented a detailed study of the gravitino LSP and Dark Matter scenario within the phenomenological
MSSM and contrasted it with the solutions available in the CMSSM. Cosmological bounds on the scenario
result in upper bounds on the gravitino mass and the reheating temperature necessary to fulfil the requirement
of a DM energy density in agreement with CMB data.

Our analysis focused on the large gravitino mass region, at
the boundary of the nucleosynthesis constraints, because that part of the parameter space allows for the largest
possible values of the reheating temperature. The present and projected LHC constraints considered in this study
include the monojet bounds that are important to test the parameter regions with large mass degeneracy, in particular
the case of higgsino and wino NLSP. The use of the unconstrained pMSSM model with 19+1 free parameters has scenarios
with bino-, wino- and higgsino-like NLSP, which are not available to studies carried out in the constrained MSSM
with 5 parameters which yield almost uniquely solutions with bino-like NLSP neutralinos.
We find that the scenario with gravitino LSP dark matter in the pMSSM is characterised by a very different composition of
the neutralino NLSP in comparison to the case of a neutralino LSP. In particular all three types of compositions are nearly
equivalently possible in the gravitino DM scenario, while for a WIMP neutralino a higgsino composition is preferred. Such a 
specific composition is unfortunately more difficult to test at a collider, since the DM mass scale tends to be above 1 TeV.
On the other hand, direct detection experiments could give a future detection, even for large neutralino mass, and
cover a large part of the yet allowed parameter space. 
But if no signal appears in searches at those experiments and in direct production at the LHC, strong 
constraints on the neutralino WIMP scenario will be obtained, making the gravitino LSP scenarios particularly interesting. 
We have seen that in case of gravitino DM and neutralino NLSP, the absence of a direct detection signal should allow
to exclude the neutralino DM scenario and point to an unstable NLSP in a large part of the parameter space. 

The interplay of the LHC and cosmology data can be pursued even further in gravitino
dark matter scenarios. The relation between the gravitino relic density and the gluino or gaugino masses, provides
us with an efficient way to test also the cosmological scenario. The region of the pMSSM parameter space with large gluino
masses, beyond the projected LHC sensitivity, corresponds to gravitino energy densities that are too large at the high reheating 
temperatures required by thermal leptogenesis. Therefore, if the next runs of the LHC and especially the High-Luminosity operation
will not observe any signal of gluino direct production, the high reheating temperature region will be excluded for
gravitino masses below 200~GeV.

\acknowledgments

AA, LC and FM acknowledge partial support from the European Union FP7 ITN INVISIBLES 
(Marie Curie Actions, PITN-GA-2011-289442), JH from the German Academy of Science through
the Leopoldina Fellowship Programme under grant LPDS 2012-14.

\end{document}